\documentstyle{article}\begin{document}
\title{Quantum state evolution in an environment of cosmological perturbations}

\author{ Z. Haba\\
Institute of Theoretical Physics, University of Wroclaw,\\
50-204 Wroclaw, Plac Maxa Borna 9, Poland\\
email:zbigniew.haba@uwr.edu.plsqueezed quantum states;stochastic equations} \maketitle
\begin{abstract}We study  the pure and thermal states of quantized scalar and tensor perturbations
in various epochs of Universe evolution.
 We calculate the density matrix of non-relativistic particles in
 an environment of these perturbations. We show that particle's
 motion can be described by a stochastic equation with a noise
 coming from the cosmological environment. We investigate the
 squeezing of Gaussian wave packets in different epochs and its
 impact upon the noise of quantized cosmological perturbations.
\end{abstract}

\section{Introduction}
The study of a system of particles with gravitational interaction
is a standard task in an investigation of inhomogeneities and
structure formation \cite{bert1}\cite{bert2}. In such studies
usually only classical gravity is considered. However, the
structure formation begins already in the inflationary era
\cite{star1}\cite{guth}\cite{hawking}\cite{chibisov}. Recent
discovery of gravitational waves raises hopes for a detection of
various phenomena resulting from quantization of gravity
\cite{bing}\cite{guer}. In the standard model of the Universe
evolution it is assumed that it begins from a quantum state. The
particles created at the end of the inflationary era will evolve
in an environment of quantized cosmological perturbations. Hence,
formation of inhomogeneities in the form of matter will take
palace in the environment of quantized perturbations.
  We can observe  the cosmological gravitational perturbations in CMB
temperature fluctuations and (possibly) in primordial
gravitational waves.  The quantum fluctuations are described in a
gauge invariant way by (gauge invariant) Bardeen scalar and tensor
variables
\cite{bardeen}\cite{mukhanov}\cite{mukhanovbook}\cite{sakai}. The
scalar variable in the inflation era is dominated by the inflaton
field. At the end of inflation the inflaton decays into
relativistic particles. The radiation era begins. We assume that
in the radiation era the quantum state of the Universe still
depends on the scalar and tensor modes of the gravitational field.
Moreover, owing to the squeezing during inflation
\cite{grishchuk}\cite{aa}\cite{star}\cite{starpol} it can be
described by a Gaussian wave function. Gaussian states are
classical in the sense that their Wigner function is positive
definite. We assume that the wave function of tensor perturbations
in spite of the complex processes taking place in various epochs
evolves in a continuous way depending only on the evolution of the
scale factor. The wave function of the scalar perturbations is not
expected to be continuous in different epochs but we still work
with a Gaussian approximation as  it is a consequence of the
quadratic approximation to Einstein gravity. The decay of the
inflaton creates particles which are moving in the environment of
the cosmological perturbations. Such an environment is changing
evolution of these particles. In a non-relativistic approximation
we derive the time evolution of the density matrix. We show that
this time evolution is determined by a stochastic equation which
is a generalization of the equation derived in
refs.\cite{wilczek}\cite{wilczek1}\cite{wilczek2}\cite{habagrav}\cite{soda}
for tensor perturbations (gravitational waves) in the Minkowski
metric. There was earlier work on the particle motion in an
environment of a quantized metric
\cite{hu}\cite{ango}\cite{habarel}\cite{hk}\cite{angohu}
\cite{wang} based on the geodesic equation. However, the
experience with the motion of a particle in a gravitational wave
\cite{wilczek}\cite{soda} indicates that the proper approach
consists in a study of microscopic quantum effects of relative
particle motions near their geodesics through the geodesic
deviation equation.

   The tensor perturbations arrive us as primordial gravitational waves.
   The scalar perturbations are measurable as temperature fluctuations in CMB
 \cite{bert2}  and as density fluctuations of galaxies \cite{bert1} \cite{bao}.
 We assume that the detector can receive primordial
perturbations from the inflationary stage of the Universe
evolution (possibly as gravitational waves produced as the second
order effect from scalar perturbations
\cite{tensor1}\cite{tensor2}). After the radiation era and
baryonic era the cosmological perturbations arrive us at the time
interval when the metric can be approximated by a static
(Minkowski) metric .
 The effect of
gravitational waves can be studied by means of a stochastic
geodesic deviation equation in a weak gravitational field on a
flat background. In \cite{habagrav} we have studied the
interaction of non-relativistic particles with quantum tensor
perturbations. We argued after \cite{wilczek}(see also
\cite{allen}) that the noise from the gravitons can be observed in
the wave detector owing to the strong squeezing during inflation.
In this paper we extend the results of \cite{habagrav} to quantum
scalar and tensor perturbations in an expanding Universe. We
suggest that the quantized scalar and tensor perturbations have an
effect upon detectors of cosmological perturbations as well as
upon formation of inhomogeneities during the radiation domination
epoch. These quantum perturbations derived as quantum
modifications of the geodesic deviation equation appear in the
form of stochastic geodesic deviation equations.

 The plan of the paper is the following. In sec.2 we
introduce our method of representing the environment of
oscillators in quantum mechanics. In sec.3 we extend it to quantum
field theory. In sec.4 we discuss the scalar  perturbations in the
inflation era. In sec.5 we study the scalar perturbations after
inflation. In sec.6 we obtain Gaussian wave function for scalar
perturbation as a solution of the Schr\"odinger equation. In sec.7
the tensor perturbations and their wave function are discussed. In
sec.8 we consider the evolution of the wave function in various
cosmological epochs. In sec.9 a non-relativistic particle
interacting with cosmological perturbations is discussed. In
sec.10 we review our version of the influence functional method in
order to derive the density matrix for a particle in an
environment of quantum cosmological perturbations. We solve
stochastic equations for cosmological perturbations (needed for
the calculation of the density matrix) in sec.11. We calculate the
density matrix in a simplified model of one-mode approximation in
sec.12. General Gaussian state of the cosmological environment is
discussed in sec.13. The particle motion in thermal environment of
cosmological perturbations is obtained in sec.14. In sec.15 we
summarize our main results and point out some extensions of our
work.

Our approach is based on a quantization of the quadratic
approximation to Einstein gravity. Such an approach is justified
in a classical theory  by a linearized coupling of the
gravitational modes to the detector as confirmed by the recent
discovery of gravitational waves. Till now there are no
indications of the quantum nature of gravitational waves
(gravitons) and the relevance of extended theories of gravity (if
dark matter and dark energy are accepted). However,  recent
observations (LIGO/Virgo, Planck2015) evoke the hope to check
various methods of quantization as well as some extensions of
Einstein gravity.  The first category includes: an exponential
parameterization \cite{salam}\cite{habaarx}\cite{russ},loop
quantization \cite{ashtekar}, effective field theory
\cite{effective},asymptotically save gravity\cite{safe}. As
possible  extensions of Einstein gravity (which eventually could
avoid the introduction of dark matter and dark energy) we mention
f(R) gravity \cite{odintsov}, Brans-Dicke gravity
\cite{brans}\cite{habaarx}, non-canonical $P(X,\phi)$ and
Horndeski gravity \cite{horndeski}\cite{fer}. These extensions are
particularly interesting in view of the possible measurement of
the difference of the light velocity and gravitational waves
velocity \cite{velocity}. We shall discuss these questions in the
last section.
\section{Feynman integral on an oscillatory background} Let us consider
first a simple model of the Schr\"odinger equation of quantum
mechanics in one dimension with the potential
$\frac{m\omega(t)^{2}x^{2}}{2}$ perturbed by a time-dependent
potential $V_{t}$
\begin{equation}
i\hbar\partial_{t}\psi_{t}=(-\frac{\hbar^{2}}{2m}\nabla_{x}^{2}+\frac{m\omega(t)^{2}x^{2}}{2}+V_{t}(x))\psi_{t}.
\end{equation}
Let  $\psi_{t}^{g}$ be a solution of  the Schr\"odinger equation
with an oscillator potential
\begin{equation}
i\hbar\partial_{t}\psi_{t}^{g}=(-\frac{\hbar^{2}}{2m}\nabla_{x}^{2}+\frac{m\omega(t)^{2}x^{2}}{2})\psi_{t}^{g}.
\end{equation}
Let us write the solution of eq.(1) in the form
\begin{equation}
\psi_{t}=\psi^{t}_{g}\chi_{t}.
\end{equation}
Inserting $\chi_{t}$ from eq.(3) into eqs.(1)-(2) we find that
$\chi_{t}$ satisfies the equation
\begin{equation}\begin{array}{l}
\partial_{t}\chi_{t}=\frac{i\hbar}{2m}\nabla_{x}^{2}\chi_{t}+\frac{i\hbar}{m}
(\nabla_{x}\ln\psi^{t}_{g})\nabla_{x}\chi_{t}-\frac{i}{\hbar}V_{t}\chi_{t}
\end{array}\end{equation}
with the initial condition
\begin{equation}
\chi_{0}=\psi_{0}(\psi_{g}^{0})^{-1}
\end{equation}
expressed by the initial conditions for $\psi_{t}$ and
$\psi^{t}_{g}$.

Eq.(4) can be considered as the diffusion equation with the
imaginary diffusion constant $\frac{i\hbar}{m}$, a time-dependent
drift $\frac{i\hbar}{m} \nabla_{x}\ln\psi^{t}_{g}$ and the
potential (killing rate)$\frac{i}{\hbar}V_{t}$.

The solution of eq.(4) is determined by the solution of the
Langevin equation
\begin{equation}
dq_{s}=\frac{i\hbar}{m}\nabla\ln\psi^{t-s}_{g}(q_{s})ds+\sqrt{\frac{i\hbar}{m}}db_{s}.
\end{equation}
Here, the Brownian motion $b_{s}$ is defined as the Gaussian
process with the covariance
\begin{equation}
E[b_{t}b_{s}]=min(t,s).\end{equation} The solution of eq.(4) is
expressed \cite{freidlin}\cite{simon} by the Feynman-Kac formula
\begin{equation}
\chi_{t}(x)=E\Big[\exp\Big(-\frac{i}{\hbar}\int_{0}^{t}dsV_{t-s}(q_{s}(x))\Big)\chi_{0}(q_{t}(x))\Big],
\end{equation}
where $q_{t}(x)$ is the solution of eq.(6) with the initial
condition $q_{0}(x)=x$ and the expectation value is over the paths
of the Brownian motion.

  A derivation of the Feynman integral (8) has been discussed earlier
 in \cite{habajpa}\cite{hababook}. An extension
 of the real diffusion processes \cite{freidlin} to a complex domain with an application to the Feynman integral
 is studied in
 \cite{doss1}\cite{alb}\cite{doss2}.

As the simplest case we consider the ground state solution of
eq.(2) (with a constant $\omega$)
\begin{equation}
\psi_{g}(x)=\Big(\frac{\pi
\hbar}{m\omega}\Big)^{-\frac{1}{4}}\exp(-\frac{m\omega}{2\hbar}x^{2}).
\end{equation}
The stochastic equation (6) reads
\begin{equation}
dq=-i\omega qdt+\sqrt{\frac{i\hbar}{m}} db.
\end{equation}
 A simple calculation  gives
\begin{equation}
\int dx\vert \psi^{g}(x)\vert^{2}E[q_{t}(x)q_{t^{\prime}}(x)]
=\frac{\hbar}{2m\omega}\exp(-i\omega\vert t-t^{\prime}\vert).
\end{equation}
The rhs of eq.(11) is the expectation value  of the time-ordered
product of Heisenberg picture position operators in the ground
state of the harmonic oscillator.
\section{Quantum field
theory} We consider the canonical field theory of a scalar
massless field with the Hamiltonian (we set the velocity of light
$c=1$)
\begin{equation}H=\frac{1}{2}\int d{\bf x}\Big( \Pi^{2}+(\nabla
\phi)^{2}+v(t)\phi^{2}\Big)+\int d{\bf x}V_{t}(\phi),
\end{equation}where $\Pi({\bf x})$ is the canonical momentum
($v(t)$ is a certain function which will be specified later),
\begin{equation}
[\phi({\bf x}),\Pi({\bf y})]=i\hbar \delta({\bf x}-{\bf y}).
\end{equation}
We solve the Schr\"odinger equation
\begin{equation}
i\hbar\partial_{t}\Psi=H\Psi.
\end{equation}

Let
\begin{equation}
\Psi_{t}=\psi_{g}^{t}\chi_{t},
\end{equation}
where $\psi_{g}^{t}$ is the solution of the Schr\"odinger equation
for free field theory
\begin{equation}
i\hbar\partial_{t}\psi_{g}^{t}=\frac{1}{2}\int d{\bf x}\Big(
\Pi^{2}+(\nabla \phi)^{2}+v(t)\phi^{2}\Big)\psi_{g}^{t}.
\end{equation}
 Then, $\chi$ satisfies the equation (an infinite
dimensional version of eq.(6))
\begin{equation} \hbar\partial_{t}\chi=\int d{\bf
x}\Big(
-\frac{i}{2}\Pi^{2}-i(\Pi\ln\psi^{t}_{g}))\Pi-iV_{t}(\phi)\Big)\chi,
\end{equation}where
\begin{equation}
\Pi({\bf x})=-i\hbar\frac{\delta}{\delta \phi({\bf x})}.
\end{equation}
 It
follows from eq.(8)that the solution of eq.(17) can be expressed
as
\begin{equation}
\chi_{t}(\phi)=E\Big[\exp\Big(-\frac{i}{\hbar}\int_{0}^{t}ds
V_{t-s}(\phi_{s})\Big)\chi_{0}\Big(\phi_{t}(\phi)\Big)\Big],
\end{equation}
where $\phi_{s}(\phi)$ is the solution of the stochastic equation
\begin{equation}
d\phi_{s}({\bf x})=i\hbar\frac{\delta}{\delta \phi({\bf
x})}\ln\psi_{g}^{t-s}ds+\sqrt{i\hbar}dW_{s}({\bf x})
\end{equation}with the initial condition $\phi$. $E[...]$ denotes an expectation value with respect to the
Wiener process (Brownian motion) defined by the covariance
\begin{equation}
E\Big[W_{t}({\bf x})W_{s}({\bf y})\Big]=min(t,s)\delta({\bf
x}-{\bf y}).
\end{equation}

 Let us consider the simplest
example:the free field. Then, the ground state is
\begin{equation}
\psi_{g}=\Big(\det(\frac{\pi}{\omega})\Big)^{-\frac{1}{4}}\exp(-\frac{1}{2\hbar}\phi\omega\phi).
\end{equation}where
\begin{displaymath}
 \omega=\sqrt{-\triangle}
 \end{displaymath}Eq.(20) reads

\begin{equation}
d\phi_{t}=-i\omega\phi_{t} dt+\sqrt{i\hbar}dW.
\end{equation} The solution is
\begin{equation}
\phi_{t}(\phi)=\exp(-i\omega t )\phi+\sqrt{i\hbar}\int_{0}^{t}
\exp(-i\omega(t-s))dW_{s}.\end{equation}

For Fourier transforms  ( in other words  in the one mode
approximation ) eq.(23) reads

\begin{equation}
d\phi_{t}({\bf k})=-i\vert {\bf k}\vert\phi_{t}({\bf k})
dt+\sqrt{i\hbar}dW({\bf k}).
\end{equation}
In subsequent sections we shall use the same notation for ${\bf
x}$ and ${\bf k}$ functions. The ${\bf k}$-representation is
useful for a smooth transition from one-mode approximations  to
infinite modes.

\section{Scalar perturbations in the era of inflation}A metric
perturbation of the flat conformal metric (with the conformal time
$\tau=\int a^{-1}(t)dt$, where $t$ is the cosmic time)
\begin{displaymath}
ds^{2}=a^{2}(d\tau^{2}-d{\bf x}^{2})
\end{displaymath}
in a special gauge (conformal Newtonian gauge with no anisotropic
stress) takes the form
\begin{equation}
ds^{2}=a^{2}((1+2\psi)d\tau^{2}-((1-2\psi)\delta_{jk}+h_{jk})dx^{j}dx^{k}).
\end{equation}
We consider a single field inflaton model of inflation ( for the
formalism with multiple scalar fields see \cite{bassett}). Then,
according to
\cite{aa}\cite{mukhanov}\cite{mukhanovbook}\cite{bassett}\cite{schwarz}
the action for scalar cosmological  perturbations in the
inflationary era (in conformal time) is
\begin{equation}
S=\frac{1}{2}\int dx\Big((\varphi^{\prime})^{2}-(\nabla
\varphi)^{2}+z^{-1}z^{\prime\prime}\varphi^{2}\Big),
\end{equation}
where \begin{equation}\varphi=a\Phi\end{equation} and $\Phi $ is a
gauge invariant variable linear in the scalar metric perturbation
$\psi$ and in the inflaton perturbation.

 The Lagrangian equations of
motion following from the action (27) are
\begin{equation}
(\partial_{\tau}^{2}-\nabla^{2}-z^{-1}z^{\prime\prime})\varphi=0
\end{equation}
where  $z$ can be expressed by the scale factor $a$ \cite{schwarz}
$z=a\sqrt{\gamma}$ with
\begin{displaymath}
\gamma=1-a^{2}(\partial_{\tau}a)^{-2}\partial_{\tau}(a^{-1}\partial_{\tau}a).
\end{displaymath}
During an inflation in a scalar potential $U$ in the slow-roll
approximation \cite{bassett} we have

\begin{equation}
z^{-1}z^{\prime\prime}=(H_{c}a)^{2}(2+5\epsilon-3\eta),
\end{equation}
where $H_{c}$ is the Hubble variable in the cosmic time and
\begin{displaymath}
\epsilon=\frac{1}{16\pi G}(\frac{U^{\prime}}{U})^{2},
\end{displaymath}\begin{displaymath}
\eta=\frac{1}{8\pi G}\frac{U^{\prime\prime}}{U}.
\end{displaymath}
$z^{-1}z^{\prime\prime}$  in the approximation of an almost
exponential expansion (i.e. for small $\epsilon$ and $\eta$) is
$2\tau^{-2}$ (as $(H_{c}a)^{2}\simeq \tau^{-2}$). Hence, in this
approximation eq.(29) reads
\begin{equation}
(\partial_{\tau}^{2}+k^{2}-2\tau^{-2})\varphi=0.
\end{equation}
The Hamiltonian (12) for the action (27) is
\begin{equation}
H=\frac{1}{2}\int d{\bf x}\Big(\Pi^{2}+(\nabla
\varphi)^{2}-z^{-1}z^{\prime\prime}\varphi^{2}\Big).
\end{equation}
The model is quantized in a standard way by a realization of the
canonical commutation relations with $ \Pi({\bf x})$ defined in
eq.(18) (now $v=-z^{-1}z^{\prime\prime}$ in eq.(13)).
\section{Scalar acoustic environment}During inflation the inflaton
field is dominant in $\Phi$ (28) but when the inflation stops, the
inflaton decays and the reheating begins (radiation era). In such
a case in the gauge invariant variable $\Phi$  the scalar
perturbations of the metric become dominant. We assume that the
scalar perturbations evolve adiabatically (constant entropy)
according to the equation for the gauge invariant scalar metric
perturbations( with no anisotropic stress and with the flat
spatial background metric) \cite{mukhanov}\cite{mukhanovbook}
\begin{equation}
\frac{d^{2}}{d\tau^{2}}\Phi+3(1+c_{s}^{2}){\cal
H}\frac{d}{d\tau}\Phi+c_{s}^{2} \triangle
\Phi+(2\frac{d}{d\tau}{\cal H}+(1+3c_{s}^{2}){\cal H}^{2})\Phi=0,
\end{equation}
where ${\cal H}=a^{-1}\frac{d}{d\tau}a$, $c_{s}$ is the acoustic
velocity approximately equal $\sqrt{w}$ and $p=w\rho$ where $p$ is
the pressure and $\rho$ is the density in the energy-momentum
tensor on the rhs of Einstein equations. The effect of the decay
of the inflanton at the end of the inflation era could be
described by a modification of eq.(33) by a friction term
$\gamma\partial_{\tau}\Phi$ \cite{warmfriction}. We assume that
either $\gamma$ is negligible or eq.(33) describes the evolution
of the scalar perturbation after the decay of the inflaton.

We consider power-law expansion in a conformal time
\begin{equation}
a= C\tau^{\alpha}. \end{equation}We introduce
\begin{equation}
\Phi=\tau^{r}\phi,
\end{equation}
where \begin{equation} r=-\frac{3}{2}(1+w)\alpha,
\end{equation}
 then the Fourier transform of eq.(33) can  be expressed as
\begin{equation}
\frac{d^{2}}{d\tau^{2}}\phi+c_{s}^{2}k^{2}
\phi-\kappa\tau^{-2}\phi=0,
\end{equation}
where $\kappa\tau^{-2}=z^{-1}z^{\prime\prime}$ of eq.(29) with
\begin{equation}
\kappa=\frac{9}{4}(1+w)^{2}\alpha^{2}+\frac{1}{2}\alpha
-\frac{3}{2}w\alpha -(1+3w)\alpha^{2}.
\end{equation}
Eq.(37) is an analog of eq.(29) with $\triangle\rightarrow
c_{s}^{2}\triangle$ . The solution of eq.(37) can be expressed by
the cylinder functions $Z_{\nu}$
\begin{equation} \phi=\sqrt{k\tau}Z_{p-\frac{1}{2}}(c_{s}k\tau)
\end{equation} where
\begin{equation}
\kappa=p(p-1)
\end{equation}The Hamiltonian corresponding to eq.(37) is
\begin{equation}
H=\frac{1}{2}\int d{\bf x}\Big(\Pi^{2}+c_{s}^{2}(\nabla
\phi)^{2}-\kappa \tau^{-2}\phi^{2}\Big),
\end{equation}
i.e.,$z^{-1}z^{\prime\prime}\rightarrow \kappa \tau^{-2}$ in
eq.(32).

\section{Gaussian solution of the Schr\"odinger equation for scalar perturbations}

We look for a solution of the Schr\"odinger equation (14) (with
the Hamiltonian (32) or (41)) in the form \begin{equation}
\psi_{t}^{g}=N\exp\Big(\frac{i}{2\hbar}\phi\Gamma_{\phi}(\tau)\phi+\frac{i}{\hbar}J_{\tau}\phi\Big),
\end{equation}where $\Gamma_{\phi}$ (we denote
$\Gamma$ in the scalar case with an index $\phi$ in order to
distinguish it from the one for the tensor perturbations in the
next sections; we shall skip $\phi$ if there is no danger of
confusion) is an operator defined by a bilinear form
$\Gamma_{\phi}({\bf x}-{\bf y})$. Inserting $\psi_{t}^{g}$ in the
Schr\"odinger equation (14) with the Hamiltonian (32) we obtain
equations for $N,\Gamma,J$ (in Fourier space)
\begin{equation}\begin{array}{l}
i\hbar\partial_{\tau}\ln N=\frac{1}{2}\int d{\bf k} J({\bf
k})J(-{\bf k})-\frac{i\hbar}{2}\delta({\bf 0})\int d{\bf
k}\Gamma({\bf k}),
\end{array}\end{equation}
where $\Gamma({\bf k})$ is the Fourier transform of $\Gamma({\bf
x})$,
\begin{equation}
\begin{array}{l}
\partial_{\tau}J=-\Gamma J,
\end{array}\end{equation}
\begin{equation}
\partial_{\tau}\Gamma=-\Gamma^{2}-c_{s}^{2}k^{2} +z^{-1}z^{\prime\prime}.
\end{equation}The term $\delta({\bf 0})$ in the normalization
factor(43)
 results from an infinite sum of oscillator energies. It could be made finite by
 a regularization of the Hamiltonian (32) but this is irrelevant for calculations of the expectation values
 (because the normalization factor cancels).
If we define
\begin{equation}
u(\tau)=\exp( \int^{\tau}ds \Gamma_{s}),
\end{equation}
then $\Gamma({\bf k})=u^{-1}\partial_{\tau}u$ where $u$ satisfies
the equation
\begin{equation}
(\partial_{\tau}^{2}+c_{s}^{2}k^{2}-
z^{-1}z^{\prime\prime})u(k)=0.
\end{equation}
With the result (30) in the inflation era ($c_{s}=1$) this
equation reads
\begin{equation}
(\partial_{\tau}^{2}+k^{2}- (2+5\epsilon-3\eta)\tau^{-2})u(k)=0.
\end{equation}
The solution is
\begin{equation}
u=C_{1}\sqrt{k\tau}Z_{p_{1}-\frac{1}{2}}(k\tau)+C_{2}\sqrt{k\tau}Z_{p_{2}-\frac{1}{2}}(k\tau),
\end{equation} where $p_{1}$ and $p_{2}$ are the solutions of the
quadratic equation
 \begin{displaymath}
p(p-1)=2+5\epsilon-3\eta\end{displaymath} and  $Z_{\nu}$ are the
cylinder functions. The solutions (49) enter the formula for the
free field  quantization in the Heisenberg picture with
 the Bunch-Davis vacuum and in the formula for the spectrum of scalar
 perturbations \cite{bunch}\cite{wise}.

For the acoustic perturbations (33) we have the Hamiltonian (41).
 The Gaussian
wave function (42) is the solution of  the Schr\"odinger equation
(14) if $\Gamma=u^{-1}\partial_{s}u$ where
\begin{equation}
\frac{d^{2}}{d\tau^{2}}u+c_{s}^{2}k^{2} u-\kappa\tau^{-2}u=0.
\end{equation}
In the era of  radiation domination ($\alpha=1$) we insert
$w=\frac{1}{3}$
 for relativistic particles, then eq.(50) reads\begin{equation}
 \frac{d^{2}}{d\tau^{2}}\phi+\frac{1}{3}k^{2}\phi
 -2\tau^{-2}\phi=0.
 \end{equation}
The Hamiltonian is
\begin{equation}
H=\frac{1}{2}\int d{\bf x}\Big(\Pi^{2}+\frac{1}{3}(\nabla
\phi)^{2}-2\tau^{-2}\phi^{2}\Big).
\end{equation}
The equation for $u$ determining $\Gamma$ reads

\begin{equation}
(\partial_{\tau}^{2}+\frac{1}{3}k^{2}- 2\tau^{-2})u(k)=0.
\end{equation}
\section{Schr\"odinger wave function for tensor perturbations}
The quadratic action for tensor (transverse, traceless)
perturbations is \cite{mukhanov} (where $\tau$ is the
conformal time)
\begin{equation}
S=\frac{1}{2}\int d\tau d{\bf
x}a^{2}(\partial_{\tau}h_{ij}\partial_{\tau}h_{ij}-\nabla
h_{ij}\nabla h_{ij}).
\end{equation}
Let us decompose $h_{ij}$ in polarization tensors $e^{\nu}_{ij}$
\begin{equation}
h_{ij}=a^{-1}e^{\nu}_{ij}h_{\nu},
\end{equation}
then
\begin{equation}
S=\int d\tau d{\bf
x}\Big(\partial_{\tau}h^{\nu}\partial_{\tau}h^{\nu}+h^{\nu}(\triangle+a^{\prime\prime}a^{-1}
 )h^{\nu}\Big).
\end{equation}
The Hamiltonian is
\begin{equation}H=\frac{1}{2}\int d{\bf
x}\Big((\Pi^{\nu})^{2}-h^{\nu}(\triangle+a^{\prime\prime}a^{-1}
 )h^{\nu}\Big),
\end{equation}
where $\Pi^{\nu}$ is the canonical momentum. After quantization
\begin{equation}
H=\frac{1}{2}\int d{\bf x}\Big(-\hbar^{2}\frac{\delta^{2}}{\delta
h^{\nu}({\bf x})^{2}}+ h^{\nu}(-\triangle-a^{\prime\prime}a^{-1}
 )h^{\nu}\Big).
\end{equation}
The Schr\"odinger equation
\begin{displaymath}i\hbar\partial_{\tau}\Psi=H\Psi
\end{displaymath}
has a Gaussian solution (where $\Gamma_{h}$ is an integral
operator with the kernel $\Gamma_{h}({\bf x}-{\bf y})$)
\begin{equation}
\psi_{\tau}^{g}=N(\tau)\exp(\frac{i}{2\hbar}h^{\nu}\Gamma_{h}(\tau)h^{\nu})
\end{equation}
if (we skip the index $h$)\begin{equation}
\partial_{\tau}\Gamma+\Gamma^{2}+(k^{2}-a^{\prime\prime}a^{-1}
 )\Gamma=0\end{equation} and
 \begin{equation}
 \partial_{\tau}\ln N=-\frac{1}{2}\delta({\bf 0})\int d{\bf k}\Gamma({\bf k}).\end{equation}
 Note that if $\Gamma$ is a continuous function of $\tau$ then
 $N(\tau)$  (hence also $\psi_{\tau}^{g}$) is a continuous
 function of $\tau$.  As we show in the next section $\Gamma$ can be continuous between different epochs of the expansion
 but the derivative of $\Gamma$
has a discontinuity between the inflationary, radiation and
baryonic epochs.
 Let as in the scalar case
 \begin{equation}
 \Gamma=u^{-1}\partial_{\tau}u
 \end{equation}
Then
\begin{equation}
\partial_{\tau}^{2}u +(k^{2}-a^{\prime\prime}a^{-1}.
 )u=0\end{equation}On the boundaries of various epochs the
Schr\"odinger equation needs some correction terms (barriers)
because  of the discontinuity of $a^{\prime\prime}$.
 In the de Sitter space (inflation era) $a=\exp(H_{c}t)$ (in the cosmic time $t$),
$\tau=-H_{c}^{-1}\exp(-H_{c}t)$ and $a=-(H_{c}\tau)^{-1}$ (in
conformal time, where $H_{c}$ is the Hubble constant in the cosmic
time) then

\begin{equation}
\partial_{\tau}^{2}u +(k^{2}-2\tau^{-2} )u=0.\end{equation}
From Friedmann equations if $p=w\rho$ (where $\rho$ is the density
and $p$ the pressure), then $ a\simeq t^{\frac{2}{3(1+w)}}$. In
the radiation era ,$w=\frac{1}{3}$, then $ a\simeq \tau$, hence in
eq.(63)
\begin{equation}
\partial_{\tau}^{2}u +k^{2}
 u=0
\end{equation}
with the general solution
\begin{equation}
u_{r}(\tau)=k\sigma_{r}\cos(k\tau)+k\delta_{r}\sin(k\tau).
\end{equation}
We consider further on mostly $\sigma\neq 0$ and $\delta\neq 0$.
We may let $\delta=0$ but then in order to obtain a normalizable
Gaussian wave packet we must shift the argument of cosine  by a
complex number $\alpha-i\gamma$. Then, as in \cite{guth}
($u=k\cos(k\tau+\alpha-i\gamma)$),
\begin{equation}\begin{array}{l}
\Gamma=-k\tan(k\tau+\alpha-i\gamma)= k\Big(i\sinh(2\gamma)-
\sin(k\tau+\alpha)\cos(k\tau+\alpha)\Big)\cr\times\Big(\cosh(2\gamma)
+\frac{1}{2}\cos(2k\tau+2\alpha)\Big)^{-1}.
\end{array}\end{equation}
Note that $\Im(\Gamma)\simeq 2k\gamma$ for a small $\gamma$
(squeezing). This form of $\Gamma$ is useful if we wish to
represent the squeezing as explicitly proportional to $\gamma$
(for a small $\gamma$) .

In the baryonic era when $w=0$ ("dust") then $a\simeq \tau^{2}$,
hence again $a^{\prime\prime}a^{-1}=2\tau^{-2}$. So, we have the
same equation (64) as for the exponential (inflationary) expansion
(when we use the approximation $\epsilon=\eta=0$).

  The solution of eq.(64)(inflationary era)
is ($ p_{1}=2,p_{2}=-1$ in eq.(49))
\begin{equation}
u=\tau^{-1}\Big(\sigma
(k\tau\cos(k\tau)-\sin(k\tau))+\delta(k\tau\sin(k\tau)+\cos(k\tau))\Big)\end{equation}
It can be seen that for the solution (66) as well as (68) $\Gamma$
depends only on $R=\delta \sigma^{-1}$. We can obtain normalizable
solutions of the Schr\"odinger equation  in the inflationary era
with $\delta=0$ shifting the arguments of $\sin$ and $\cos$ by a
complex factor. So instead of (68) we can write a solution of
eq.(64) in the form
\begin{equation}
u=k\cos(k\tau+\alpha-i\gamma)-\tau^{-1}\sin(k\tau+\alpha-i\gamma).
\end{equation}
The solution of the acoustic equations (33)(37) and (50) is
obtained from eqs.(68)-(69) with $k\tau\rightarrow c_{s}k\tau$. So
eq.(53) (describing scalar perturbations in the radiation era) for
an acoustic wave of a relativistic fluid has a solution analogous
to eq.(68)
\begin{equation}\begin{array}{l}
u=\tau^{-1}\Big(\sigma
(k\tau\cos(\frac{1}{\sqrt{3}}k\tau)-\sin(\frac{1}{\sqrt{3}}k\tau))
+\delta(k\tau\sin(\frac{1}{\sqrt{3}}k\tau)+\cos(\frac{1}{\sqrt{3}}k\tau))\Big).\end{array}\end{equation}
An analog of eq.(69) is

\begin{equation}
u=k\cos(\frac{1}{\sqrt{3}}k\tau+\alpha-i\gamma)-\tau^{-1}\sin(\frac{1}{\sqrt{3}}k\tau+\alpha-i\gamma).
\end{equation}
These solutions can be used in sec.13 for a calculation of the
density matrix in the environment of scalar perturbations in the
radiation era.

We view the time evolution of the Gaussian wave function of tensor
perturbations as a continuous process through various epochs of
Universe evolution. We wish to follow the Gaussian wave function
starting from the inflationary era. For this purpose we need to
choose  the expansion scale $a(\tau)$ in a continuous way. We use
\cite{grishchuk}
\begin{displaymath} a(\tau)=\tau^{-1}K\tau_{1}^{-1}
\end{displaymath}
in the inflationary era when $-\infty<\tau<\tau_{1}<0$
\begin{displaymath}
a(\tau)=-\tau_{1}^{-2}(\tau -2\tau_{1})K\tau_{1}^{-1}
\end{displaymath}
in the radiation era when $\tau_{1}<\tau<\tau_{2}$ and
\begin{equation}
a(\tau)=-\frac{1}{4}(\tau+\tau_{2}-4\tau_{1})^{2}\tau_{1}^{-2}(\tau_{2}-2\tau_{1})^{-1}K\tau_{1}^{-1}
\end{equation} for $\tau>\tau_{2}$ ( baryonic era).

$a(\tau)$ is continuous together with its first derivative. Hence,
$\Gamma(\tau)$ can be glued together in a continuous way (so that
the wave function is continuous). $\frac{d^{2}a}{d\tau^{2}}$ is
discontinuous (does not exist at the transition points between
different eras). In such a case eq.(63) (equivalent to the
Schr\"odinger equation (14)) requires an interpretation. Eq.(63)
is like the Schr\"odinger equation in one-dimension with a
discontinuous potential $a^{-1}a^{\prime\prime}$. When
$a^{-1}a^{\prime\prime}$ is discontinuous  between  different
epochs then  we must impose continuity conditions upon $u_{s}$ as
in the Schr\"odinger equation on the line with discontinuous
barriers.
\section{Evolution of the quantum Gaussian state in various cosmological epochs}
In this section we investigate whether squeezing of the wave
function for scalar and tensor perturbations
\cite{grishchuk}\cite{aa}\cite{star} $\Im i\Gamma\simeq -\tau^{2}$
(at small $\tau$)achieved in the inflation era continues to the
subsequent epochs of the Universe evolution assuming that the
Gaussian wave function is continuous between different epochs (it
satisfies the Schr\"odinger equation with the quadratic
Hamiltonians in the corresponding epochs). The squeezing is
relevant for the noise intensity in the equations of motion of
quantum particles \cite{wilczek} as will be shown in sec.13. We
suggest that the expansion of the Universe itself ( not the
physical processes in various epochs) has the major impact on the
wave function evolution at least for the tensor perturbations.
This can be justified by the fact that in the first order
approximation the gravitational waves created
 during inflation  interact neither with the scalar perturbations
nor with the matter created  after inflation. For the scalar
perturbations this assumption may be questionable because the
inflanton decays between inflation and radiation era so that in
the radiation era its contribution to $\Phi$ of eq.(33) is
diminishing. Finally, only the scalar metric perturbation remains
in $\Phi$. However in the second order perturbative calculations
the scalar perturbations can produce the gravitational waves
\cite{tensor1}\cite{tensor2}. Such non-linear effects cannot be
described by a Gaussian approximation. The quantum state varies
with $\Gamma (\tau) $ where $\Gamma (\tau) $ is determined by $u$.
We begin the evolution of the state $\psi_{\tau}(h)$ in the
inflationary era. Then, eqs.(59),(62)-(64) apply. We assume that
when the inflation stops at $\tau_{1}$ then the radiation era
begins. Then, the evolution of $\psi_{\tau}(h)$ is determined by
eq.(66). After the recombination at $\tau_{2}$ the photons
decouple. We assume that in this era $w=0$. Then, again eq.(64) is
satisfied (but now $a\simeq \tau^{2}$).

  For a continuous evolution of the wave function
$\psi_{\tau}(h)$  we need the continuity conditions for
$\Gamma(\tau)$ at the start $\tau_{1}$ of the radiation  era and
at the beginning $\tau_{2}$ of the baryonic era. Let us denote
$R=\sigma^{-1}\delta$ ( the functions $\Gamma$ are defined by $R$.
During the inflation we have (from eq.(68))\begin{equation}
\begin{array}{l}
\Gamma_{i}(\tau)=\Big(-k^{2}\sin(k\tau)-k\tau^{-1}\cos(k\tau)+\tau^{-2}\sin(k\tau)\cr
+R_{i}k^{2}\cos(k\tau)-R_{i}k\tau^{-1}\sin(k\tau)-R_{i}\tau^{-2}\cos(k\tau)\Big)\times
\cr\Big(k\cos(k\tau)-\tau^{-1}\sin(k\tau)+R_{i}k\sin(k\tau)+R_{i}\tau^{-1}\cos(k\tau)\Big)^{-1}
\end{array}
\end{equation}
For small $k\tau$ (large cosmic time $t$) we obtain
\begin{equation}
\Gamma_{i}(\tau)=\tau^{-1}\Big(-1+(k\tau)^{2}-\frac{1}{R_{i}}(k\tau)^{3}\Big)
\end{equation}
From eq.(74) the real part of $i\Gamma_{i}$ is small (squeezing
 \cite{aa}) whereas the imaginary part of $i\Gamma_{i}$ is large
(classical WKB behaviour\cite{aa}). The assumption that at
$\tau\rightarrow -\infty$ the state $\psi^{g}_{\tau}$ tends to the
vacuum requires $R_{i}=i$. For $R_{i}=i$
 we have from eq.(73) exactly\begin{equation}
\Gamma_{i}=ik\Big(1+i(k\tau)^{-3}\Big)\Big(1+(k\tau)^{-2}\Big)^{-1}
\end{equation}
which  agrees with the approximate formula (74).

In the begin of the radiation era at time $\tau_{1}$ (from
eq.(66))
\begin{equation}\begin{array}{l}
\Gamma_{r}(\tau_{1})=\Big(-k^{2}\sin(k\tau_{1})+k^{2}R_{r}\cos(k\tau_{1})\Big)
 \Big(k\cos(k\tau_{1})+kR_{r}\sin(k\tau_{1})\Big)^{-1}
\end{array}\end{equation}

If the wave function $\psi^{g}_{\tau}(h)$ is to be continuous
between the inflation era and the radiation era then from
$\Gamma_{i}(\tau_{1})=\Gamma_{r}(\tau_{1})$ we get
\begin{equation}
R_{r}=-\Big(k\sin(k\tau_{1})+\Gamma_{i}\cos(k\tau_{1})\Big)\Big(\Gamma_{i}\sin(k\tau_{1})-k\cos(k\tau_{1})\Big)^{-1}
\end{equation}
 and subsequently we can express $R_{r}$ by $R_{i}$

\begin{equation}\begin{array}{l}
R_{r}=\Big(k\tau_{1}-\sin(k\tau_{1})\cos(k\tau_{1})-R_{i}(k^{2}\tau_{1}^{2}-\cos^{2}(k\tau_{1})\Big)\cr\times
\Big(-k^{2}\tau_{1}^{2}+\sin^{2}(k\tau_{1})-k\tau_{1}R_{i}-\sin(k\tau_{1})\cos(k\tau_{1})R_{i}\Big)^{-1}
\end{array}\end{equation}

At small $k\tau_{1}$ this gives
\begin{equation}
R_{r}\simeq -\frac{1}{2k\tau_{1}}+\frac{i}{4}(k\tau_{1})^{2}
\end{equation}

 If
$k\tau_{1}$ is large then from eq.(78)
\begin{equation}
R_{r}\simeq R_{i}
\end{equation}
$\kappa\tau_{1}$ is large in the baryonic era. As we shall show in
sec.13 (see also \cite{habagrav}) large $R$ ensures big noise.In
order to have a large noise in the radiation era (and subsequently
in the baryonic era) we need big $R_{i}$ in the inflation era.

As in the inflation era and in the baryonic era the same eq.(64)
applies we have the same continuity conditions in the passage
between different eras (eq.(72)). Eq.(77) remains true in the
baryonic era (for $\tau>\tau_{2}$) when we replace $R_{i}$ by $
R_{b}$ (the squeeze parameter in the baryonic era) but now $\tau$
is large. It follows from eq.(80) that $R_{b}\simeq R_{r }$.

\section{Particle interacting with scalar and
tensor perturbations in the post-inflationary era} We are
interested in the motion of a particle in the gravitational field
of cosmological perturbations. The particles ("baryons" or dark
matter particles) appear at the end of the inflation era. The
environment of the cosmological perturbations may have an impact
upon the clustering (formation of inhomogeneities) of such
particles. The environment of tensor perturbations ( quantized
gravitational waves) may be detected  in LIGO/Virgo detector as
suggested in \cite{wilczek}. This is a system of mirrors such that
the interference takes place depending on the distance of the
mirrors. Let us consider  a system of two particles (mirrors) with
masses $m^{\prime}>>m$ in a free falling frame (e.g. in satellites
as in prospective LISA gravitational wave detector) . The distance
between the particles depends on the metric
$ds^{2}=g_{\mu\nu}dX^{\mu}dX^{\nu}$ . The action for the light (m)
particle is
\begin{displaymath}S=-m\int \sqrt{g_{\mu\nu}dX^{\mu}dX^{\nu}}
\end{displaymath} where the metric is defined in eq.(26).
If we are to compute averages over the metric we need to expand
the action in  a perturbation series in weak gravitational
perturbations. In order to do it in a covariant way we choose the
Fermi coordinates between the space-like separated geodesics of
the particles $m^{\prime}$ and $m$. Now, $X^{\mu}=(\tau,{\bf X})$,
where ${\bf X}$ are the Fermi coordinates between the neighboring
geodesics of the two particles. Calculating the square root in $S$
in the lowest order with the metric (26) we obtain
\cite{wilczek}\cite{soda}
\begin{displaymath}
S=\frac{m}{2}\int d\tau a(\frac{d{\bf X}}{d\tau})^{2}
-\frac{m}{2}\int d\tau a R_{0l0r}X^{r}X^{l},
\end{displaymath}
where $R_{\mu\nu\alpha\beta}$ is the Riemannian tensor. Inserting
the expression for the Riemannian tensor with a linear
approximation (27) for the metric we obtain the action

\begin{equation}\begin{array}{l}
S_{I}=\frac{m}{2}\int d\tau
a\frac{dX^{k}}{d\tau}\frac{dX^{k}}{d\tau} -\lambda m\int d\tau
\psi(\tau,{\bf X}(\tau))\frac{d^{2}}{d\tau^{2}}a{\bf X}^{2}
\cr+\frac{1}{2}m\lambda\int d\tau h_{jk}(\tau,{\bf
X}(t))\frac{d^{2}}{d\tau^{2}}(aX^{k}X^{j}),
\end{array}\end{equation}
where $\lambda^{2}=8\pi G$ ( we have rescaled $\psi\rightarrow
\lambda\psi,h\rightarrow \lambda h$ so that the quadratic Einstein
gravitational action for the  perturbations is the same as the one
for the free massless scalar field of sec.3).

In a general metric perturbation the action should depend on gauge
invariant variables. Hence, $\psi\rightarrow \Phi$ in
eq.(81).According to eq.(35)
$\phi=\tau^{\frac{3}{2}(1+w)\alpha}\Phi$  so the action (81) takes
the form
\begin{equation}\begin{array}{l}
S_{I}=\frac{m}{2}\int d\tau
a\frac{dX^{k}}{d\tau}\frac{dX^{k}}{d\tau} -\lambda m\int d\tau
\tau^{-\frac{3}{2}(1+w)\alpha}\phi(\tau,{\bf
X}(\tau))\frac{d^{2}}{d\tau^{2}}a{\bf X}^{2}
\cr+\frac{1}{2}m\lambda\int dt h_{jk}(\tau,{\bf
X}(\tau))\frac{d^{2}}{d\tau^{2}}(aX^{k}X^{j}),
\end{array}\end{equation}
where $\phi$ satisfies eq.(37) and its quantum evolution is
determined by the Hamiltonian (41)( in the radiation era
$w=\frac{1}{3}$, $\alpha=1$ , $\phi=\tau^{2}\Phi$ ). The tensor
field $h_{ij}$ is expressed by $h^{\nu}$ (55) with the Hamiltonian
(57)

 The interaction of a particle with the
gravitational waves in eq.(82) is the same (for $a=1$) as the one
in \cite{wilczek}\cite{soda} (the scalar terms have been
considered in \cite{zurek}). Eq.(82) defines a linear coupling
model with the coupling
\begin{equation}
V=\int d{\bf x}(\phi f+h_{rl}f^{rl}),
\end{equation}
where
\begin{equation}
f({\bf x})=-m\lambda\int d\tau\tau^{-\frac{3}{2}(1+w)\alpha}
\delta({\bf x}-{\bf X}(\tau))  \frac{d^{2}}{d\tau^{2}}a{\bf X}^{2}
\end{equation} and
\begin{equation}
f^{rl}({\bf x})=\frac{1}{2}m\lambda\int d\tau \phi\delta({\bf
x}-{\bf X}(\tau)) \frac{d^{2}}{d\tau^{2}}\Big(
aX^{r}(\tau)X^{l}(\tau)\Big).
\end{equation}
 We apply the formula for the
linear coupling in order to calculate the density matrix of
quantum particles.

From the action (82) neglecting the dependence of fields on ${\bf
X}$ (as we shall do in subsequent sections for the quantized
fields) we obtain an equation of motion for a particle in a
classical solution $\psi^{cl}$ and $h_{jk}^{cl}$ of the Einstein
equations for the gravitational perturbations
\begin{equation}
\frac{d}{d\tau}a\frac{dX_{k}}{d\tau}+\lambda\frac{d^{2}h^{cl}_{jk}}{d\tau^2}aX^{j}
+2\lambda\frac{d^{2}\psi^{cl}}{d\tau^2}aX_{k}=0.\end{equation}

\section{Linear coupling to an oscillatory environment}
We are interested in quantum mechanics of particles interacting
with quantized cosmological perturbations.  The quantized tensor
perturbations are expected to be detectable \cite{wilczek} as
gravitons. The quantum scalar perturbations are analogs of phonons
in solid state physics or plasmons in the physics of plasma. We
expect that they have an impact on detectors of cosmological
perturbations (e.g. that they can be measurable as the noise as
suggested in \cite{wilczek} for tensor perturbations). We
approximate the non-linear Einstein action by its quadratic term
in scalar and tensor perturbations. For both the scalar
perturbations (84) as well as the tensor perturbations (85) the
interaction with a particle  takes place through a linear coupling
of an oscillatory system with the coordinates of the particle. In
one-mode approximation and with a linearized interaction there are
no substantial differences between the scalar and the tensor
terms. We consider a model of a system with a Lagrangian ${\cal
L}_{X}$ described by a coordinate $X$ interacting  linearly with
an oscillator $q$ ($q$ can be either the scalar or the tensor
mode). We denote the particle part of the interaction by
$f_{s}(X)$.
  We have the Lagrangian
\begin{equation}
{\cal L}={\cal L}_{X}+
\frac{1}{2}((\frac{dq}{ds})^{2}-\omega^{2}q^{2})+ q_{s}f_{s}(X).
\end{equation}
We could quantize the interaction (87) in the Heisenberg picture
solving Lagrange equations with quantized oscillators $q$. Such an
approach for an analogous electromagnetic interaction has been
developed in \cite{benguria}\cite{ford}. One obtains a quantum
system ${{\cal L}_{X}}$ with a noise expressed by the oscillator
creation and annihilation operators. Similar treatment of the
particle-gravity interaction is discussed in \cite{soda}. Then,
the noise does not depend on the quantum state of the oscillators
but the correlation functions of the noise must be calculated in a
particular quantum state of these oscillators.

In the approach of sec.2 the noise depends on the state of the
oscillatory background. When we calculate the expectation values
of the observables of the $X$-system (which are independent of the
$q$-variables) then according to quantum mechanics the calculation
is reduced to an evaluation of the trace in the mixed state
$\rho_{t}$
\begin{equation}\begin{array}{l}
Tr_{q}(_{t}\vert \Phi><\Phi\vert _{t})\equiv \rho_{t}
\end{array}\end{equation}
where $Tr_{q}$ is the trace over the states of the $q$- subsystem
and  $\vert \Phi>$ is the pure state  of the system (87). We
consider an initial state of the product form
$\Phi(q,X)=\psi_{0}^{g}(q)\chi_{0}(q)\phi(X)$. According to eq.(3)
it evolves into $\psi_{t}^{g}(q)\chi_{t}(q)\phi_{t}(q,X)$ where
\begin{displaymath}\chi_{t}(q)\phi_{t}(q,X)
=E[\chi_{0}(q_{t}(q))\phi(q_{t}(q),X)]\end{displaymath} where in
$E[\chi_{0}(q_{t}(q))\phi(q_{t},X)]$ the evolution of the
$\chi_{0}\phi$ state is expressed by the ordinary Feynman integral
over the $X$ paths and the expectation value over oscillator paths
of sec.2 (so $q_{t}(q) $ is the stochastic process (10)). When the
initial state of the oscillator is fixed as $\psi_{0}^{g}(q)$ (as
in secs.(12)-(13)) then the average (88) is reduced to a q-average
over $\vert \psi_{t}^{g}\vert^{2}$. For the thermal state of
sec.14 the average (88) will be over all states of the oscillator
with a proper Gibbs weight.

In this paper we consider the system of particles and cosmological
fluctuations. We  do no measurements on gravitational
fluctuations. Nevertheless, these fluctuations have some impact
upon the motion of quantum particles. As will be shown in the
following sections the effect of the  fluctuations upon the
particle's motion can be described classically as a friction and
noise. The averaging applies also to classical fields (including
classical gravitational fluctuations). It would not make much
difference whether we derived the density matrix considering,
e.g., classical background of gravitational waves or coherent
states of quantized gravitational waves. In the Appendix of
\cite{habagrav} we have shown that the assumption that
gravitational waves have classical thermal distribution leads to
the same friction and noise as the high temperature limit in the
average (88) over the quantum Gibbs distribution. Nevertheless,
for lower temperature the noise and friction are
$\hbar$-dependent. The effect of a classical cosmological
background is weak (proportional to the Newton constant) whereas
we can expect a strong detectable noise from quantum squeezed
states as discussed in sec.13 .

Our  approach  can be considered as a tool for a calculation of
Feynman-Vernon influence functional \cite{feynman}. Using the
stochastic representation of sec.2 the density matrix $\rho$ of
the system ${\cal L}_{X}$ is obtained as an average over the
environment of the oscillator in the state $ \psi_{t}^{g}$
\begin{equation}\begin{array}{l}
\rho_{t}(X,X^{\prime})=\int dx {\cal D}X{\cal
D}X^{\prime}\vert\psi_{t}^{g}(x)\vert^{2}\exp(-\frac{i}{\hbar}\int
ds {\cal L}_{X}+\frac{i}{\hbar}\int ds {\cal
L}_{X^{\prime}})\overline{\phi_{i}}(X_{t}(X))\phi_{i}(X_{t}(X^{\prime}))\cr
E\Big[\exp\Big(\frac{i}{\hbar}\int_{0}^{t}q_{s}f_{t-s}(X_{s}^{\prime})\Big)\chi_{i}(q_{t}(x))
\exp\Big(-\frac{i}{\hbar}\int_{0}^{t}q^{*}_{s}f_{t-s}(X_{s})\Big)\overline{\chi_{i}}(q^{*}_{t}(x))\Big].
\end{array}\end{equation} where ${\cal D}X$ means the Feynman integral over the particle's trajectories,
${\cal D}X^{\prime}$ is an integral over independent Feynman paths
$X^{\prime}$ and
 $*$ when acting on functions means the
complex conjugation and when applied to the stochastic process
(10) it means a complex conjugation of an independent version of
the process (10).

 For a Gaussian variable $q_{s}$ we have (for any
number $\alpha_{s}$)
\begin{displaymath}\begin{array}{l}
E[\exp(\alpha_{s}q_{s})]=\exp\Big(\alpha_{s}<q_{s}>+\frac{1}{2}<(\alpha_{s}q_{s}-\alpha_{s}<q_{s}>)^{2}>\Big).
\end{array} \end{displaymath} This equation can easily be generalized to
$\int ds\alpha_{s}q_{s}$ . If $\chi_{i}=1$ the expectation value
in eq.(89) is
\begin{equation}\begin{array}{l}
E\Big[\exp\Big(\frac{i}{\hbar}\int_{0}^{t}dsq_{s}f_{t-s}(X_{s})\Big)\Big]
=\exp\Big(\frac{i}{\hbar}\int_{0}^{t}ds<q_{s}>f_{t-s}(X_{s})\Big)
\cr\exp\Big(-\frac{1}{2\hbar^{2}}\int_{0}^{t}dsds^{\prime}
E[(q_{s}-<q_{s}>)(q_{s^{\prime}}-<q_{s^{\prime}}>)]
f_{t-s}(X_{s})f_{t-s^{\prime}}(X_{s^{\prime}})\Big).\end{array}\end{equation}

\section{Solution of the stochastic equation for scalar and tensor perturbations}
According to the results of sec.2 and sec.10 the calculation of
the Feynman integral is reduced to the calculation of expectation
values over  solutions of stochastic equations. For the scalar
perturbation with $\psi_{g}^{t}$ of eq.(42) eq.(20) has the
solution (with the initial condition $\phi$ at $\tau=\tau_{0}$)
\begin{equation}\begin{array}{l}
\phi_{s}=\frac{u_{\tau-s}}{u_{\tau-\tau_{0}}}\phi
-u_{\tau_{0}}u_{\tau-s}J_{0}\int_{\tau_{0}}^{s}u_{\tau-t}^{-2}dt+
\sqrt{i\hbar}u_{\tau-s}\int_{\tau_{0}}^{s}u_{\tau-t}^{-1}dW_{t},
\end{array}
\end{equation} where $u_{s}$ is a solution of eq.(47) (in subsequent sections we denote
by the same  symbol  solutions of eq.(47) which are different
depending on the choice of $c_{s}$ and $z^{-1}z^{\prime\prime}$ in
eq.(47)). We have
\begin{equation}\begin{array}{l}
E[(\phi_{s}-<\phi_{s}>)(\phi_{s^{\prime}}-<\phi_{s^{\prime}}>)]
=i\hbar
u_{\tau-s}u_{\tau-s^{\prime}}\int_{\tau_{0}}^{min(s,s^{\prime})}u_{t-\tau}^{-2}d\tau.\end{array}
\end{equation}
With  $V$ linear in $\phi$ in eq.(19)  we can calculate the
expectation value in eq.(90) explicitly using  eqs. (91)-(92).

Linearized gravity decomposed in polarization components $h^{\nu}$
(55) has the  Hamiltonian (57) which is  the same as the one for
two independent scalar fields. Hence, the solution of the
Schr\"odinger equation  of the linearized Einstein gravity is the
product of the solutions for the scalar fields $h_{\nu}$ (a
generalization of eq.(59) with a source term $J$ which can
describe coherent states of the gravitational waves; we assume
that owing to the rotation invariance $\Gamma_{h}$ does not depend
on $\nu$ )\begin{equation}
\psi_{g}^{\tau}(h)=A(\tau)\exp\Big(\frac{i}{2\hbar}(h^{\nu}\Gamma_{h}(\tau)h^{\nu}+2J^{\nu}_{\tau}h_{\nu})\Big),
\end{equation}where
$\Gamma$ is an integral operator with the kernel $\Gamma(\tau,{\bf
x}-{\bf y})$. As in sec.6 we find that the Fourier transform of
$\Gamma(\tau,{\bf x}-{\bf y})$ can be expressed as
$\Gamma(\tau,{\bf k})=u({\bf k})^{-1}\partial_{\tau}u({\bf k})$
and $J_{\tau}=J_{0}u_{0}u_{\tau}^{-1}$. Then, Eq.(20)
  takes the form
\begin{displaymath}
dh_{s}^{\nu}=-\Gamma(\tau-s)h_{s}^{\nu}ds-J_{\tau-s}^{\nu}ds+\sqrt{i\hbar}dW^{\nu}_{s},
\end{displaymath}where
\begin{displaymath}
E[W_{t}^{\alpha}({\bf x})W_{s}^{\beta}({\bf
y})]=min(t,s)\delta^{\alpha\beta}\delta({\bf x}-{\bf y}).
\end{displaymath}Expressing $\Gamma$ and $J$ in terms of $u$ (eq.(62)) we have
(we suppress the index $\nu$)
\begin{equation}\begin{array}{l} dh_{s}=-\partial_{\tau}\ln
u_{\tau-s}
h_{s}ds-J_{0}u_{0}u_{\tau-s}^{-1}ds+\sqrt{i\hbar}dW_{s}.
\end{array}
\end{equation}
Eq.(94) has the solution (with the initial condition $h$ at
$\tau=\tau_{0}$)
\begin{equation}\begin{array}{l}
h_{s}(h)=\frac{u_{\tau-s}}{u_{\tau-\tau_{0}}}h
-u_{\tau_{0}}u_{\tau-s}J_{0}\int_{\tau_{0}}^{s}u_{\tau-t}^{-2}dt+
\sqrt{i\hbar}u_{\tau-s}\int_{\tau_{0}}^{s}u_{\tau-t}^{-1}dW_{t}.
\end{array}
\end{equation}
We have
\begin{equation}\begin{array}{l}
E[(h_{s}^{\alpha}-<h_{s}^{\alpha}>)(h_{s^{\prime}\beta}-<h_{s^{\prime}\beta}>)]
=i\hbar\delta^{\alpha\beta}
u_{\tau-s}u_{\tau-s^{\prime}}\int_{\tau_{0}}^{min(s,s^{\prime})}u_{t-\tau}^{-2}d\tau,\end{array}
\end{equation}
 where  for the solution (66) ($R=\sigma^{-1}\delta$)
\begin{equation}
\int_{0}^{s}u_{t-\tau}^{-2}d\tau= k^{-4}(1+R^{2})^{-1}(\Gamma_{h}(
t-s)-\Gamma_{h}(t))
\end{equation}
with \begin{equation}
\Gamma_{h}(s)=k\Big(-\sin(ks)+R\cos(ks)\Big)\Big(\cos(ks)+R\sin(ks)\Big)^{-1}
\end{equation}
whereas for the solution (68) \begin{equation}\begin{array}{l}
\sigma k^{3}\int_{t}^{t^{\prime}}u_{\tau}^{-2}d\tau=
(kt^{\prime}\sin(kt^{\prime})+\cos(kt^{\prime}))\Big((\delta
kt^{\prime}-\sigma)\sin(kt^{\prime})+(\sigma
kt^{\prime}+\delta)\cos(kt^{\prime})\Big)^{-1}\cr-(kt\sin(kt)+\cos(kt))\Big((\delta
kt-\sigma)\sin(kt)+(\sigma kt+\delta)\cos(kt)\Big)^{-1}\equiv
\sigma k^{-1}(\Gamma(t)-\Gamma(t^{\prime}))
\end{array}
\end{equation}
 We use the notation $\Gamma$ at the rhs of eq.(99) in order to comply with the formulas for the correlation
 functions of $\phi_{s}$ in eqs.(92) and (97). For the scalar perturbations in the radiation era when we have the
solution $u_{s}$ (of eq.(70)) still the result (99) applies with
$k\rightarrow c_{s}k$ with $c_{s}=\frac{1}{\sqrt{3}}$.

 If $\delta=0$ and
$u=k\cos(k\tau+\alpha-i\gamma)$ then
\begin{equation}\begin{array}{l}
\int_{t}^{t^{\prime}}u_{s}^{-2}ds =k^{-3}\Big(
\tan(kt^{\prime}+\alpha-i\gamma)- \tan(kt+\alpha-i\gamma)\Big)  =
k^{-4}(\Gamma_{h}(t)-\Gamma_{h}(t^{\prime})),\end{array}\end{equation}
where $\Gamma$ is defined in eq.(67). In the analogous formula in
the inflation era
\begin{equation}\begin{array}{l}
k^{3}\int_{t}^{t^{\prime}}u_{\tau}^{-2}d\tau\cr
=\Big(kt^{\prime}\sin(kt^{\prime}+\alpha-i\gamma)+\cos(kt^{\prime}+\alpha-i\gamma)\Big)
\Big(-\sin(kt^{\prime}+\alpha-i\gamma)+kt^{\prime}\cos(kt^{\prime}+\alpha-i\gamma)\Big)^{-1}
\cr-\Big(kt\sin(kt+\alpha-i\gamma)+\cos(kt+\alpha-i\gamma)\Big)
\Big(-\sin(kt+\alpha-i\gamma)+kt\cos(kt+\alpha-i\gamma)\Big)^{-1}
\cr \equiv
k(\Gamma^{i}_{h}(t)-\Gamma^{i}_{h}(t^{\prime}))\end{array}\end{equation}
The formulas (97) and (99)-101) allow to calculate the  evolution
of the density matrix (88) for the interaction of particles with
the cosmological perturbations.
\section{One mode approximation} In our linearized model (87) $q$ will be
either $\phi$ or $h_{jk}$ and $f$ is defined in eqs.(84)-(85).In
coordinate space (with an infinite number of modes) if $\psi_{g}$
is the ground state (22) then we have
\begin{displaymath} \int d\phi\vert
\psi_{g}(\phi)\vert^{2}\exp(iF\phi)
=\exp(-\frac{\hbar}{4}F\omega^{-1}F)
\end{displaymath}
for any function $F$ and
\begin{equation} \omega^{-1}({\bf x},{\bf y}) =D ({\bf
x},{\bf y})=(2\pi)^{-3}\int d{\bf k}k^{-1}\exp(i{\bf k}({\bf
x}-{\bf y}))
\end{equation} For
the scalar perturbation we set $q_{s}$ as
\begin{equation}
\phi_{s}({\bf k})=\exp(-iks)\phi({\bf
k})+\sqrt{i\hbar}\int_{0}^{s}\exp(-ik(s-\tau))dW_{\tau}({\bf k})
\end{equation}
In order to exhibit the method without an involvement with
cumbersome formulas we perform the functional integration for one
mode first (repeating for the convenience of the reader some
calculations of \cite{habagrav}). We begin with the simplest case
of the background of one single oscillator in the ground state
(9). The average of one mode $ \phi({\bf k})$ in eq.(89) is
calculated as
\begin{equation}\begin{array}{l}
\int d\phi({\bf k})\exp\Big(-\frac{1}{\hbar}\phi({\bf
k})^{*}k\phi({\bf k}) \Big)\exp(i\phi({\bf k})^{*}F({\bf
k})+iF({\bf k})\phi^{*}({\bf k})))\cr =\exp (-\hbar F({\bf
k})^{*}k^{-1}F({\bf k})).\end{array}
\end{equation}
We  calculate the expectation value (89) for the density matrix of
the $X$-system assuming that the oscillator is in the ground state
and we do not calculate expectation values of any  oscillator
observables. According to eqs.(90)-(92) we obtain (we use the
solution (103), assume that the initial condition $\chi_{i}=1$ and
we take only one component of ${\bf X}$)
\begin{equation}\begin{array}{l}\rho_{\tau}\simeq\int dx{\cal D}X{\cal D}X^{\prime}\cr\exp(-\frac{k x^{2}}{\hbar})
 E\Big[\exp\Big(\frac{i}{\hbar}\int_{0}^{\tau}(
\frac{m}{2}\frac{dX}{ds}\frac{dX}{ds}-\frac{m}{2} \frac{dX^{\prime
}}{ds}\frac{dX^{\prime }}{ds}-\frac{i}{\hbar}
\int_{0}^{\tau}(q_{\tau-s}f_{s}-q^{*}_{\tau-s}f^{\prime}_{s})ds\Big)\Big]\cr
=\int dx{\cal D}X{\cal
D}X^{\prime}\exp\Big(\frac{i}{\hbar}\int_{0}^{\tau}(
\frac{m}{2}\frac{dX}{ds}\frac{dX}{ds}-\frac{m}{2} \frac{dX^{\prime
}}{ds}\frac{dX^{\prime }}{ds}\cr\exp(-\frac{k x^{2}}{\hbar})
\exp\Big(-\frac{i}{\hbar} \int_{0}^{\tau}(x\exp(-ik
(\tau-s))f_{s}-x\exp(i\omega (\tau-s))f^{\prime}_{s})ds\Big)
\cr\exp\Big(-\frac{1}{2\hbar^{2}}\int_{0}^{\tau}dsds^{\prime}

\Big(E[(q_{\tau-s}-<q_{\tau-s}>)(q_{\tau-s^{\prime}}-<q_{\tau-s^{\prime}}>)]f_{s}f_{s^{\prime}}\cr
+E[(q^{*}_{\tau-s}-<q^{*}_{\tau-s}>)(q^{*}_{\tau-s^{\prime}}-<q^{*}_{\tau-s^{\prime}}>)]
f^{\prime}_{s}f^{\prime}_{s^{\prime}}\Big)\Big),
\end{array}
\end{equation} here $f_{s}^{\prime}=f_{s}(X^{\prime})$. In eq.(105) we have (this is the special case of
eq.(92) with $u_{s}=\exp(ik s)$)
\begin{equation}\begin{array}{l}
E[(q_{\tau-s}-<q_{\tau-s}>)(q_{\tau-s^{\prime}}-<q_{\tau-s^{\prime}}>)]\cr=\frac{\hbar}{2k}\Big(\exp(-ik\vert
s-s^{\prime}\vert) -\exp(-ik(2\tau-s-s^{\prime}))\Big).\end{array}
\end{equation}
If the oscillator is in a time-dependent state then  we should
insert the solution (66) (or (69)) in the Feynman formula (88).
Hence, instead of eq.(105) we have
\begin{equation}
\begin{array}{l}\int dx\vert\exp(i\frac{\Gamma(\tau)
x^{2}}{2\hbar})\vert^{2} E\Big[\exp\Big(\frac{i}{\hbar}
\int_{0}^{\tau}(q_{s}f_{\tau-s}-q^{*}_{s}f^{\prime}_{\tau-s})ds\Big)\Big]\cr
=\int dx\exp(i\frac{\Gamma(\tau)
x^{2}}{2\hbar})\exp(-i\frac{\Gamma^{*}(\tau) x^{2}}{2\hbar})
 \exp\Big(\frac{-i}{\hbar}
\int_{0}^{\tau}(<q_{\tau-s}>f_{s}-<q_{\tau-s}^{*}>f^{\prime}_{s})ds\Big)
\cr\exp\Big(-\frac{1}{2\hbar^{2}}\int_{0}^{\tau} dsds^{\prime}
\Big(E[(q_{\tau-s}-<q_{\tau-s}>)(q_{\tau-s^{\prime}}-<q_{\tau-s^{\prime}}>)]f_{s}f_{s^{\prime}}\cr
+E[(q^{*}_{\tau-s}-<q^{*}_{\tau-s}>)(q^{*}_{\tau-s^{\prime}}-<q^{*}_{\tau-s^{\prime}}>)]
f^{\prime}_{s}f^{\prime}_{s^{\prime}}\Big)\Big),
\end{array}
\end{equation}
where
\begin{equation}\begin{array}{l}
E[(q_{\tau-s}-<q_{\tau-s}>)(q_{\tau-s^{\prime}}-<q_{\tau-s^{\prime}}>)]=i\hbar
u_{s}u_{s^{\prime}}\int_{0}^{min(\tau-s,\tau-s^{\prime})}dt
u(\tau-t)^{-2}\cr=-i\hbar
k^{-2}u_{s}u_{s^{\prime}}(\sigma^{2}+\delta^{2})^{-1}(\Gamma(\tau)-\Gamma(max(s,s^{\prime}))).
\end{array}\end{equation}
In our model $q_{s}$ is $\phi_{s}$ for the scalar perturbation and
\begin{equation}
f_{s}({\bf k})=-m\lambda s^{-\frac{3}{4}(1+w)\alpha}\exp(i{\bf
k}{\bf X}_{s})\frac{d^{2}}{ds^{2}}a{\bf X}^{2}
\end{equation}
In the tensorial case $q_{s}$ is $h_{s}^{rl}$ and
\begin{equation}
f_{s}^{rl}({\bf k})=\frac{m}{2}\lambda\exp(i{\bf k}{\bf
X}_{s})\frac{d^{2}}{ds^{2}}aX^{r}X^{l}
\end{equation}

The calculation of the $x$ integral in eqs.(105) and (107) leads
to a quadratic functional of $f_{s}$ and $f^{\prime}_{s^{\prime}}$
in the exponential. In the  case (105) of the ground state of the
one dimensional oscillator the  scalar  fluctuations give (now
$a=1$ and $\alpha=0$)\begin{equation}\begin{array}{l}
\rho_{\tau}\simeq \exp\Big(\frac{i}{\hbar}\int_{0}^{\tau}\Big(
\frac{m}{2} \frac{dX}{ds}\frac{dX}{ds}-\frac{m}{2}
\frac{dX^{\prime }}{ds}\frac{dX^{\prime }}{ds}\Big)\cr
\exp\Big(-\frac{1}{4\hbar k}\int_{0}^{t}
dsds^{\prime}\Big(f_{s}f_{s^{\prime}}\exp(-ik\vert
s-s^{\prime}\vert)+f^{\prime}_{s}f^{\prime}_{s^{\prime}}\exp(ik\vert
s-s^{\prime}\vert)-2f_{s^{\prime}}f^{\prime}_{s}\exp(ik(s-s^{\prime}))\Big)\Big).
\end{array}\end{equation}
We write \begin{equation} Q=\frac{1}{2}(X+X^{\prime })
\end{equation}
\begin{displaymath}
y=X-X^{\prime }.
\end{displaymath}We expand the exponential in eq.(111) in $y$. We obtain \begin{equation}\begin{array}{l}\rho_{t}=
\int{\cal D}Q{\cal D}y
\exp\Big(\frac{i}{\hbar}\int_{0}^{\tau}ym\frac{d^{2}Q}{ds^{2}}\Big)\cr
\exp\Big(-\frac{1}{\hbar k}\int_{0}^{\tau}
ds^{\prime}\int_{0}^{s^{\prime}}ds\Big(-i\sin(k(s-s^{\prime}))(\frac{d^{2}}{ds^{2}
}yQ\frac{d^{2}}{ds^{\prime 2 }}Q^{2}
\cr-\frac{d^{2}}{ds^{2}}yQ\frac{d^{2}}{ds^{\prime
2}}yQ\cos(k(s-s^{\prime}))\Big)
\rho_{0}(Q_{\tau},y_{\tau})\end{array}\end{equation} The term
linear in $y$ modifies the equation of motion of the $Q$
coordinate. The term quadratic in $y$ is a noise acting upon the
particle \cite{hk}.

In the expression (107) of the time-dependent reference state we
obtain
\begin{equation}\begin{array}{l} \rho_{\tau}\simeq
\exp\Big(-\frac{i}{2\hbar}(\Gamma(\tau)-\Gamma^{*}(\tau))^{-1}\Big(\int_{0}^{\tau}(u_{\tau}^{-1}u_{s}f_{s}
 -u_{\tau}^{*-1}u_{s}^{*}f_{s}^{\prime})ds\Big)^{2}\cr
 -\frac{i}{2\hbar k^{2}}
 \int_{0}^{\tau}ds\int_{0}^{s^{\prime}}ds^{\prime}\Big(u_{s}u_{s^{\prime}}(\sigma^{2}+\delta^{2})^{-1}
(\Gamma(\tau)-\Gamma(max(s,s^{\prime})))f_{s}f_{s^{\prime}}\cr
-u^{*}_{s}u^{*}_{s^{\prime}}(\sigma^{* 2}+\delta^{* 2})^{-1}
(\Gamma^{*}(\tau)-\Gamma^{*}(max(s,s^{\prime})))f^{\prime}_{s}f^{\prime}_{s^{\prime}}
\Big)dsds^{\prime}\Big).\end{array}
\end{equation}
We  expand (114) in $y$ . After the expansion  in $y$ till the
second order terms in eq.(105) we obtain
\begin{displaymath}\begin{array}{l}
\int {\cal D}Q {\cal D}y
\exp\Big(\frac{i}{\hbar}\int_{0}^{\tau}y(m\frac{d^{2}Q}{ds^{2}}
+L(Q)+\frac{i}{2\hbar}My)\Big)\rho_{0}(Q,y)\cr=\int {\cal D}Q
{\cal
D}y\exp\Big(\frac{i}{\hbar}\int_{0}^{\tau}(y\tilde{L}+\frac{i}{2\hbar}yMy)\Big)\rho_{0}(Q,y)\cr
=\int {\cal D}Q {\cal D}y \exp\Big(-\frac{1}{2\hbar^{2}}(y-i\hbar
M^{-1}\tilde{L})M(y-i\hbar M^{-1}\tilde{L})
-\frac{1}{2}\tilde{L}M^{-1}\tilde{L}\Big)
\rho_{0}(Q,y),\end{array}\end{displaymath}where by $L$ we denote a
functional of $Q$, $M$ is an operator and by $y\tilde{L}$ we
denote the term proportional to $y$. We introduce $\tilde{Q}=
M^{-\frac{1}{2}}\tilde{L}$, then $\tilde{Q}$ is a Gaussian
variable which has the white noise distribution that can be
represented as $\partial_{s}b_{s}$. It can be seen that the
equation $\tilde{Q}= M^{-\frac{1}{2}}\tilde{L}$ can be expressed
as the stochastic equation
\begin{equation}
m\frac{d^{2}Q}{ds^{2}}+L(Q)=M^{\frac{1}{2}}\partial_{s}b_{s}.
\end{equation}The calculation of
$\rho_{\tau}$ involves an average over solutions of the stochastic
equation (115). In general, there still will be the Gaussian
integral over $y$ so that the expression for the density matrix
can be obtained in the form of an integral over the solutions of
the stochastic equation (115) and over the $y$ terms resulting
from an expansion in $y$ of $\rho_{0}(Q,y)$ (this is an expansion
in $\hbar$).

\section{General Gaussian environment of  cosmological perturbations}
In this section we consider a general Gaussian time-dependent
state  of scalar and tensor perturbations. These perturbations are
generated by independent scalar fields $\phi, h^{\nu}$.  The
difference between scalar and tensor perturbations is in the way
they couple to particle velocities ( eqs.(109)-(110)).The action
(82) (together with the gravitational action) in Fourier space
takes the form of a sum over ${\bf k}$ modes ( in the interaction
of a particle with cosmological perturbations we neglect the
dependence of $h(s,{\bf x})$ on spatial coordinates )
\begin{equation}
\begin{array}{l}
S=\int ds\Big(\frac{1}{2}\int d{\bf k} h^{*}_{\nu}(s,{\bf
k})(-\partial_{s}^{2}-k^{2}-a^{-1}a^{\prime\prime})h_{\nu}(s,{\bf
k})\cr+ \int d{\bf k} \phi(s,{\bf
k})(-\partial_{s}^{2}-c_{s}^{2}k^{2}-\kappa s^{-2})\phi(s,{\bf
k})\cr +\frac{1}{2}ma\frac{dX_{r}}{ds}\frac{dX_{r}}{ds}
+(2\pi)^{-\frac{3}{2}}\int d{\bf k} h_{\nu}(s,{\bf
k})f_{s}^{\nu}({\bf k}) +(2\pi)^{-\frac{3}{2}}\int d{\bf
k}\phi_{s}({\bf k})f_{s}({\bf k})\Big)\Big),
\end{array}\end{equation}
where $h_{\nu}=a^{-1}e^{\nu}_{rl}h^{rl}$, $f^{\nu}({\bf
k})=e^{\nu}_{rl}f^{rl}({\bf k})$ and $f^{rl}({\bf k}),f({\bf k})$
are defined in eqs.(109)-(110).

 We consider a solution of the Schr\"odinger equation in an expanding universe which has
the Gaussian form
\begin{equation}
\psi^{g}_{\tau}(h,\phi)=\exp\Big(\frac{i}{2\hbar}\int d{\bf
k}h^{\alpha}\Gamma_{h}(\tau) h^{\alpha}+\frac{i}{2\hbar}\int d{\bf
k}\phi\Gamma_{\phi}(\tau) \phi\Big).
\end{equation}
As discussed in sec.8 in an expanding universe $\Gamma(\tau)$ in
$\psi^{g}_{\tau}$ can dramatically change in time so that  $\Im
i\Gamma\simeq -\tau^{2}$ (squeezing)
\cite{grishchuk}\cite{aa}\cite{star}. We show in this section that
calculating expectation values in terms of the density matrix
(obtained by averaging over $\vert \psi^{g}_{\tau}\vert^{2}$)
according to eqs.(88)-(89) leads to a large noise on the basis of
eq.(115).

 The calculation of the
functional integral with the action (116) in an environment of
cosmological perturbations (117) is reduced (according to sec.10)
to a calculation of expectation values with respect to the
stochastic processes $\phi_{s}$ and $h^{\nu}_{s}$. These
stochastic processes and their correlation functions are defined
by the solutions $u_{s}$ of eq.(47) with various $c_{s}$ (for the
scalar perturbations) and $z^{-1}z^{\prime\prime}$. The general
result for a calculation of the expectation values is contained in
eq.(90) but the complexity of the detailed formulas depends on the
complexity of the solution $u_{s}$. In the remaining part of this
section we write down explicitly the expressions for the
environment of the  $\phi$ field which is in the ground state in
eq.(117) or in the time dependent (squeezed) state with
$z^{-1}z^{\prime\prime}=0$ (Minkowski space-time). We outline the
calculations
 for the scalar perturbations in the radiation era when
 $z^{-1}z^{\prime\prime}=2\tau^{-2}$.
For the tensor perturbations we have explicit elementary solutions
for $u_{s}$
 and $h_{s}^{\nu}$ and their correlations (eq.(99)) so that we can calculate the density
matrix in the environment of the tensor perturbations exactly in
all epochs of the universe evolution.

 The
scalar part of the contribution to the density matrix in the
non-expanding metric $ a=1$ (described by the solution (66)) or in
the radiation era (described by the solution (70) ) is
\begin{equation}\begin{array}{l} \rho_{\tau}\simeq
\exp\Big(-\frac{1}{2\hbar}\int d{\bf
k}(\Gamma(\tau)-\Gamma(\tau)^{*})^{-1}\Big(\int_{0}^{\tau}(u_{\tau}^{-1}u_{s}f_{s}
 -u_{\tau}^{*-1}u_{s}^{*}f_{s}^{\prime})ds\Big)^{2}\cr
 -\frac{i}{2\hbar}\int d{\bf k}k^{-4} \int_{0}^{\tau}\Big(u_{s}u_{s^{\prime}}(\sigma^{2}+\delta^{2})^{-1}
(\Gamma(\tau)-\Gamma(\mu)))f_{s}f_{s^{\prime}}
\cr-u^{*}_{s}u^{*}_{s^{\prime}}(\sigma^{* 2}+\delta^{* 2})^{-1}
(\Gamma^{*}(\tau)-\Gamma^{*}(\mu))f^{\prime }_{s}f^{\prime
}_{s^{\prime}}\Big)dsds^{\prime}\Big).\end{array}
\end{equation}
where $\mu=max(s,s^{\prime})$ and in the radiation era $
(\sigma^{2}+\delta^{2})^{-1}\Gamma$ should be replaced by $\Gamma$
from eq.(99), $f_{s}$ is defined in eq.(109) where we neglect
$\exp(i{\bf kX})$.

The tensor part in the expectation values (105) and (107) with an
infinite number of modes has been calculated in \cite{habagrav}
(for $a=1$). We shall have sums of the form
\begin{equation}
f_{s}^{\nu}f_{s^{\prime}}^{\nu}=\Lambda_{mn;rl}f^{mn}_{s}f^{rl}_{s^{\prime}},
\end{equation}where
$\Lambda_{rl;mn}=e^{\nu}_{rl}e^{\nu}_{mn}$. The result of an
averaging over angles  leads to an insertion $f\rightarrow q^{rl}$
in eq.(118) where\begin{equation}
q^{rl}=\frac{m}{2}a^{-1}\frac{d^{2}}{ds^{2}}(X^{r}X^{l}-\frac{2}{3}\delta^{rl}{\bf
X}^{2}\Big)a.
\end{equation}Hence, the
tensor contribution to the expectation value (118)
is\begin{equation}\begin{array}{l} \rho_{\tau}\simeq
\exp\Big(-\frac{i}{2\hbar}\int d{\bf
k}(\Gamma(\tau)-\Gamma(\tau)^{*})^{-1}\Big(\int_{\tau_{0}}^{\tau}(u_{\tau}^{-1}u_{s}f^{\alpha}_{s}
 -u_{\tau}^{*-1}u_{s}^{*}f_{s}^{\prime\alpha})ds\Big)^{2}\cr
 -\frac{i\lambda^{2}}{2\hbar}\frac{4\pi}{5}\int dk k^{-2}\int_{\tau_{0}}^{\tau}\Big(u_{s}u_{s^{\prime}}(\sigma^{2}+\delta^{2})^{-1}
\cr(\Gamma(\tau)-\Gamma(\mu))q^{rl}_{s}q^{rl}_{s^{\prime}}
-u^{*}_{s}u^{*}_{s^{\prime}}(\sigma^{* 2}+\delta^{* 2})^{-1}
(\Gamma^{*}(\tau)-\Gamma^{*}(\mu))q^{\prime rl}_{s}q^{\prime
rl}_{s^{\prime}}\Big)dsds^{\prime}\Big).\end{array}
\end{equation} where $u_{s}$ in general would be the solution of
eq.(63) but in the static metric and in the radiation era
$a^{-1}a^{\prime\prime}=0$, hence we have the solution (66). We
could use the formula (121) for tensor perturbations in the
inflationary and baryonic era when
$a^{-1}a^{\prime\prime}=2\tau^{-2}$. Then, the function $\Gamma$
is defined in eq.(99) or(101). The factor $\frac{4\pi}{5}$ in
eq.(121) comes from the average over angles in the $d{\bf k}$
integration. The behavior of $\rho_{\tau}$ depends on the complex
function $R(k)=\sigma(k)\delta(k)^{-1}$ in eq.(121). We can see
that the final noise term can be large because of the squeezing
factor $(\Gamma-\Gamma^{*})^{-1}$ in eq.(121).Explicitly, the term
with $(\Gamma-\Gamma^{*})^{-1}$ coming from the scalar
perturbation is
\begin{equation}\begin{array}{l}

\exp\Big(-\frac{i\lambda^{2}}{2\hbar}4\pi\int dk
k^{2}(\Gamma(\tau)-\Gamma(\tau)^{*})^{-1}\int_{\tau_{0}}^{\tau}dsds^
{\prime}\Big(u_{\tau}^{-2}u_{s}u_{s^{\prime}}f_{s}f_{s^{\prime}}\cr
+u_{\tau}^{*-2}u_{s}^{*}u_{s^{\prime}}^{*}f_{s}f^{\prime}_{s^{\prime}}
- u_{\tau}^{-1}u_{\tau}^{*-1}u_{s}^{*}u_{s^{\prime}}f^{\prime
}_{s}f_{s^{\prime}}
-u_{\tau}^{-1}u_{\tau}^{*-1}u_{s}u_{s^{\prime}}^{*}f_{s}f^{\prime
}_{s^{\prime}}\Big)\Big).
\end{array}
\end{equation}
For a comparison let us first calculate the expression (122) for
the ground state (22) of the scalar field. Then
$\frac{\delta}{\sigma}=i$, $\Gamma=ik$ , $u_{s}=\exp(iks)$ and
 in
 the integral (122) we obtain (this is an
 infinite mode version of eq.(111))
\begin{equation}\begin{array}{l}\rho_{\tau}\simeq
\exp\Big(-\frac{1}{2\hbar}\int d{\bf
k}k^{-1}\int_{\tau_{0}}^{\tau} dsds^{\prime}\cr
\times\Big(\exp(-ik\vert s-s^{\prime}\vert)f_(s)f_{s^{\prime}})
+\exp(ik\vert s-s^{\prime}\vert)
f^{\prime}(s)f^{\prime}(s^{\prime})\cr
-\exp(-ik(s-s^{\prime}))(f_{s}f^{\prime}_{s^{\prime}}+f^{\prime}_{s}f_{s^{\prime}})\Big).
\end{array}
\end{equation}After the expansion (112) eq.(123) gives
a term linear in $y$
\begin{equation}\begin{array}{l}
-4i\frac{\lambda^{2}}{2\hbar}\int d{\bf
k}k^{-1}\int_{\tau_{0}}^{\tau}ds^{\prime}\frac{d^{2}}{ds^{\prime
2} }Q^{r}y^{r} \int_{0}^{s^{\prime}}ds\frac{d^{2}}{ds^{2}} {\bf
Q}^{2}\sin(k(s^{\prime}-s)).
\end{array}\end{equation} Representing $\sin(ks)$ in
eq.(124) as $-k^{-1}\partial_{s}\cos(ks)$ we integrate over $k$
obtaining $\partial_{s}\delta(s-s^{\prime})$. This is a local
radiation damping term which coincides with the one which will be
obtained in sec.14 (eq.(144)) for thermal gravitational
perturbations. The noise resulting from eq.(123) can be read from
the quadratic part in eq.(123)
\begin{equation}\begin{array}{l}\rho_{\tau}\simeq
\exp\Big(-\frac{\lambda^{2}m^{2}}{8\hbar }\int d{\bf
k}k^{-1}\int_{\tau_{0}}^{\tau}
ds^{\prime}\int_{\tau-{0}}^{s^{\prime}}ds\Big(-4\frac{d^{2}}{ds^{\prime
2}}Q^{r}y^{r}
\frac{d^{2}}{ds^{2}}y^{l}Q^{l}\cos(k(s-s^{\prime}))\Big)
\rho_{0}(Q_{t},y_{t})\end{array}\end{equation} For a time
dependent $\psi_{t}^{g}$ we work out eq.(121) in more explicit
form (in the Minkowski space-time $a=1$ ) calculating

\begin{equation}
\Gamma(\tau)-\Gamma(\tau)^{*}=k^{3}(R-R^{*})(u_{\tau}u_{\tau}^{*})^{-1},
\end{equation}
\begin{equation}
\Gamma(\tau)-\Gamma(s)=k^{3}u_{\tau}^{-1}u_{s}^{-1}\sin(k(s-\tau))(1+R^{2}).
\end{equation}
Using eqs.(126)-(127) we obtain from eq.(122)
\begin{equation}\begin{array}{l}
\rho_{\tau}\simeq\exp\Big(-\frac{i}{2\hbar}\int dsds^{\prime}\int
d{\bf
k}k^{-3}\cr\times\Big((R-R^{*})^{-1}(u_{s}u_{s^{\prime}}u_{\tau}^{-1}u_{\tau}^{*}f_{s}
f_{s^{\prime}}+
u_{s}^{*}u_{s^{\prime}}^{*}u_{\tau}^{*-1}u_{\tau}f_{s}^{\prime}
f_{s^{\prime}}^{\prime} -2u_{s}u_{s^{\prime}}^{*}f_{s}
f_{s^{\prime}}^{\prime})\cr-
u_{s}u_{s^{\prime}}u_{\tau}^{-1}u_{\mu}^{-1}\sin(k(\mu-\tau))f_{s}
f_{s^{\prime}}
+u_{s}^{*}u_{s^{\prime}}^{*}u_{\tau}^{*-1}u_{\mu}^{*-1}\sin(k(\mu-\tau))f^{\prime}_{s}
f_{s^{\prime}}^{\prime}\Big)\Big),\end{array}
\end{equation}
where $\mu=max(s,s^{\prime})$.

In an expansion in ${\bf y}$ (112) of the scalar contribution the
term that modifies equations of motion (a phase factor ) is
\begin{equation}\begin{array}{l}
\exp\Big(-\frac{i}{2\hbar}\int dsds^{\prime}\int d{\bf
k}k^{-3}\cr\times\Big((R-R^{*})^{-1}
(u_{\tau}^{*}u_{s}+u_{\tau}u_{s}^{*})(u_{\tau}^{-1}u_{s^{\prime}}-u_{\tau}^{*-1}u_{s^{\prime}}^{*})\cr
-\sin(\mu-\tau))(u_{s}u_{s^{\prime}}u_{\tau}^{-1}u_{\mu}^{-1}+u_{s}^{*}u_{s^{\prime}}^{*}
u_{\tau}^{*-1}u_{\mu}^{*-1})\Big)   \frac{d^{2}}{ds^{\prime
2}}Q^{n}y^{n}\frac{d^{2}}{ds^{2}}{\bf Q}^{2}\Big)\Big).
\end{array}\end{equation}
The scalar contribution to the noise in the time dependent
environment $\psi_{\tau}^{g}$ ($a=1$)can be written as (the term
in eq.(128) quadratic in $y$)

\begin{equation}\begin{array}{l}
\exp\Big(-\frac{i}{2\hbar}\int d{\bf k}k^{-3}\int
dsds^{\prime}\cr\times \Big((R-R^{*})^{-1}
(u_{\tau}^{*}u_{s}+u_{\tau}u_{s}^{*})(u_{\tau}^{-1}u_{s^{\prime}}+u_{\tau}^{*-1}u_{s^{\prime}}^{*})\cr
-\sin(k(\mu-\tau))(u_{s}u_{s^{\prime}}u_{\tau}^{-1}u_{\mu}^{-1}-u_{s}^{*}
u_{s^{\prime}}^{*}u_{\tau}^{*-1}u_{\mu}^{*-1})\Big)
\Big(\frac{d^{2}}{ds^{\prime 2}}Q^{l}y^{l}
\frac{d^{2}}{ds^{2}}Q^{n}y^{n}\Big)\Big)\cr\equiv\exp(-\frac{1}{2\hbar^{2}}{\bf
y}M{\bf y}).
\end{array}\end{equation}
Eq.(121) simplifies if $\Gamma(\tau)\simeq const$. Set in eq.(66)
$S\delta=i\sigma $ ( $S$ may depend on $k$). Then
$\Gamma(0)=ikS^{-1}$. We have a real Gaussian function in eq.(93)
as an initial state. If $S$ is large and $(k\tau)^{-1}>>S>>k\tau$
with $k\tau<<1$ then to eq.(118) only the term (122) contributes,
where $u_{s}\simeq \cos(ks)$. The density matrix is a product of
scalar and tensor terms (we assume that
$\Gamma_{\phi}=\Gamma_{h}$)
\begin{equation}\begin{array}{l} \rho_{\tau}\simeq
\exp\Big(-\frac{1}{4\hbar}\int d{\bf
k}\frac{S}{k}\Big(\int_{\tau_{0}}^{\tau}(\cos(k\tau))^{-1}\cos(ks)(f_{s}
 -f_{s}^{\prime})ds\Big)^{2}\Big)\cr\times\exp\Big(-\frac{1}{4\hbar}\int d{\bf
k}\frac{S}{k}\Big(\int_{\tau_{0}}^{\tau}(\cos(k\tau))^{-1}\cos(ks)(f^{\alpha}_{s}
 -f_{s}^{\prime\alpha})ds\Big)^{2}\Big).\end{array}
\end{equation}
The integration over the angles  $k^{-1}{\bf k}$ of
$\epsilon_{rl}^{\alpha}\epsilon_{mn}^{\alpha}$ is expressed by
$<\Lambda_{rl;mn}>$ where $\Lambda_{rl;mn}$ is defined in eq.(119)
and
\begin{displaymath}
<\Lambda_{ij;mn}>=4\pi\Big(\frac{1}{5}(\delta_{im}\delta_{jn}+\delta_{in}\delta_{jm})
-\frac{2}{15}\delta_{ij}\delta_{nm}\Big).
\end{displaymath}
 Hence, finally when we put together the scalar and tensor terms
we obtain in eq.(131)

\begin{equation}\begin{array}{l} \rho_{\tau}\simeq
\exp\Big(-\frac{\lambda^{2}m^{2}}{16\hbar}4\pi\int dkk
S(k)\Big(\int_{\tau_{0}}^{\tau}dsds^{\prime}(\cos(k\tau))^{-2}\cos(ks)\cos(ks^{\prime})
\cr\times\Big(\frac{1}{4\pi}<\Lambda_{rl;mn}>
 \frac{d^{2}}{ds^{2}}(Q^{r}y^{l}+Q^{l}y^{r})\frac{d^{2}}{ds^{\prime 2}}(Q^{m}y^{n}+Q^{n}y^{m})
 +8\delta_{rl}\delta_{mn})\frac{d^{2}}{ds^{\prime 2}}Q^{l}y^{l}
\frac{d^{2}}{ds^{2}}Q^{n}y^{n}\Big)\Big).\end{array}
\end{equation}
Eq.(132) gives the spectrum of the noise  as $8\pi G S k$ .

It is useful to express the results (128)-(132) on the evolution
of the density matrix ( in  Minkowski background space-time) with
the representation (67) of $\Gamma$ (with
$u_{s}=k\cos(ks+\alpha-i\gamma)$). Now
\begin{equation}
\Gamma(\tau)-\Gamma^{*}(\tau)=ik\sinh(2\gamma)\Big(\cosh^{2}\gamma
-\sin^{2}(k\tau)\Big)^{-1}
\end{equation}
and
\begin{equation}\begin{array}
{l}
\Gamma(\tau)-\Gamma(\tau^{\prime})=k\sin(k(\tau^{\prime}-\tau))
\Big(\cos(k\tau+\alpha-i\gamma)
\cos(k\tau^{\prime}+\alpha-i\gamma)\Big)^{-1}
\end{array} \end{equation}Eq.(118) for the scalar perturbation reads
\begin{equation}\begin{array}{l} \rho_{\tau}\simeq
\exp\Big(\frac{1}{2\hbar}\int d{\bf k}k^{-1}
\Big(\sinh(2\gamma)(\cosh^{2}\gamma-\sin^{2}(k\tau+\alpha)\Big)^{-1}\cr\times\Big(\int_{\tau_{0}}^{\tau}
(u_{\tau}^{-1}u_{s}f_{s}
 -u_{\tau}^{*-1}u_{s}^{*}f_{s}^{\prime})ds\Big)^{2}
 \cr-\frac{i}{2\hbar}4\pi\int dk k^{-3}\int_{\tau_{0}}^{\tau}dsds^{\prime}\Big(u_{s}u_{s^{\prime}}
\sin(k\mu
-k\tau)\times\cr\Big(\cos(k\tau+\alpha-i\gamma)\cos(k\mu+\alpha-i\gamma)\Big)^{-1}f_{s}f_{s^{\prime}}
\cr-u^{*}_{s}u^{*}_{s^{\prime}}\sin(k\mu
-k\tau)\Big(\cos(k\tau+\alpha+i\gamma)\cos(k\mu+\alpha+i\gamma)\Big)^{-1}f^{\prime
}_{s}f^{\prime }_{s^{\prime}}\Big)\Big),\end{array}
\end{equation}
where  $\mu=max(s,s^{\prime})$. We have
\begin{equation}
\begin{array}{l}
u_{\tau}^{-1}u_{s}f_{s}
 -u_{\tau}^{*-1}u_{s}^{*}f_{s}^{\prime}=\lambda m\Big(\cosh^{2}\gamma
 -\sin^{2}(k\tau+\alpha)\Big)^{-1}\times \cr\Big(i\sin(k(s-\tau))\sinh (2\gamma)\frac{d^{2}}{ds^{2}}({\bf
 Q}^{2}+\frac{1}{4}{\bf y}^{2})\cr
 +(\cos(k(s-\tau))\cosh (2\gamma)+\cos(k(s+\tau)))\frac{d^{2}}{ds^{2}}{\bf
 Qy}\Big)\end{array}\end{equation}
 Hence for small $\gamma$ the  term (122) is dominating. Its
 contribution to the noise is
\begin{equation}\begin{array}{l} \rho_{\tau}\simeq
\exp\Big(-\frac{\lambda^{2}m^{2}}{2\hbar}\int\frac{ d{\bf
k}}{k}\int ds ds^{\prime}
\Big(\sinh(2\gamma)\Big)^{-1}(\cosh^{2}\gamma-\sin^{2}(k\tau+\alpha)\Big)^{-1}\cr\times
\Big((\cos(k(s-\tau))\cosh
(2\gamma)+\cos(k(s+\tau)))\frac{d^{2}}{ds^{2}}{\bf
 Qy}\cr\times(\cos(k(s^{\prime}-\tau))\cosh (2\gamma)+\cos(k(s^{\prime}+\tau))))\frac{d^{2}}{ds^{\prime 2}}{\bf
 Qy}\Big)\Big).\end{array}
\end{equation}
It follows from eq.(137) that the  noise can be large if $\gamma$
is small.

We can calculate the density matrix in the radiation era when
$a(s)=s$ with the contribution of the scalar and tensor
perturbations. We  use the results (70) and (99). We do not write
down these complicated expressions. Let us mention only the
contribution of tensor perturbations to the noise. So the
contribution of the tensor perturbations to the quadratic part of
the density matrix  in the radiation era $ (a(\tau)=\tau$) is
expressed as the following  modification of eq.(132):
\begin{equation}\begin{array}{l} \rho_{\tau}\simeq
\exp\Big(-\frac{\lambda^{2}m^{2}}{16\hbar}\int dkk
S(k)\Big(\int_{\tau_{0}}^{\tau}dsds^{\prime}(\cos(k\tau))^{-2}\cos(ks)\cos(ks^{\prime})
 ((a(s)a(s^{\prime}))^{-1}\cr\times <\Lambda_{rl;mn}>
 \frac{d^{2}}{ds^{2}}a(Q^{r}y^{l}+Q^{l}y^{r})\frac{d^{2}}{ds^{\prime 2}}a(Q^{m}y^{n}+Q^{n}y^{m})\Big).\end{array}
\end{equation}
\section{Thermal  perturbations}
 In this section we first consider the constant metric $a=1$  (Minkowski
 space-time) for scalar and tensor perturbations. In such a case the Hamiltonian (32)
 for the scalar field as well as the Hamiltonian (58)
 for radiation coincide with the Hamiltonians of free massless relativistic scalar fields.
 During the radiation era the Hamiltonian in eq.(58) for the tensor field in the conformal time
 also consists of a sum of two  scalar free massless fields (12) with $v=V=0$ (as $a^{\prime\prime}=0$). For
the scalar perturbations this does not happen unless $a=1$. We
consider first the thermal scalar perturbation on a Minkowski
space-time. The geodesic deviation in thermal gravitational waves
$h^{rl}$ has been studied in \cite{habagrav}. At the end of this
section we derive a modification of the geodesic deviation
equation  in a thermal state  in the radiation era. The Lagrangian
(27) and (83) with an infinite number of modes in the coordinate
space is
\begin{equation}
\begin{array}{l}
{\cal L}=\frac{1}{2} \int d{\bf x}\phi(s,{\bf
x})(-\partial_{s}^{2}+\triangle)\phi(s,{\bf x})+\int d{\bf x}\phi
f.
\end{array}\end{equation} The  evolution  of the density is
obtained from eq.(88) where the partial trace over the oscillator
states involves all eigenstates of energy $\epsilon_{n}$ with the
weight factor $\exp(-\beta\epsilon_{n})$ with $\beta
=\frac{1}{k_{B}T}$ where $k_{B}$ is the Boltzmann constant and $T$
is the temperature. This partial trace has been calculated in
\cite{habagrav}\cite{calchu}\cite{kleinert} with the result
\begin{equation}\begin{array}{l}
\rho_{\tau}({\bf X},{\bf X}^{\prime})=\int {\cal D}X {\cal
D}X^{\prime}\exp(\frac{im}{2\hbar}\int_{0}^{\tau}ds(\frac{dX_{r}}{ds}\frac{dX_{r}}{ds}
-\frac{dX_{r}^{\prime}}{ds}\frac{dX_{r}^{\prime}}{ds})
\cr\exp\Big(\frac{1}{\hbar^{2}}\int_{0}^{\tau}ds\int_{0}^{s}ds^{\prime}\Big((f-f^{\prime
})C(f+f^{\prime})-(f-f^{\prime })A(f-f^{\prime
})\Big)\rho_{0}({\bf X}_{\tau},{\bf
X}_{\tau}^{\prime}).\end{array}\end{equation} In (140) we have a
decomposition of the finite temperature  propagator $D$ into the
real and imaginary parts $D=A+iC$
\begin{equation}\begin{array}{l}
A({\bf x}-{\bf x}^{\prime},s-s^{\prime})=2 \hbar(2\pi)^{-3}\int
\frac{\bf d{\bf k}}{2k}\cos({\bf k}({\bf x}-{\bf x}^{\prime}))
\cos (k(s-s^{\prime}))\coth(\frac{\hbar\beta k}{2}),\end{array}
\end{equation}\begin{equation}\begin{array}{l}
C({\bf x}-{\bf x}^{\prime},s-s^{\prime})=2\hbar(2\pi)^{-3}\int
\frac{\bf d{\bf k}}{2k}\cos({\bf k}({\bf x}-{\bf x}^{\prime}))
\sin (k(s-s^{\prime})).\end{array}
\end{equation} In eqs.(141)-(142)
We  neglect the ${\bf x}$ dependence of the propagators and
average over the angles. Then, the $k$-integral $d{\bf k}\simeq
4\pi dkk^{2}$ in the high temperature limit $\beta\hbar
\rightarrow 0$ of $A$ in eq.(141) gives $\delta(s-s^{\prime})$. In
$C$ (142) we write (as in \cite{hk})
$\sin(k(s-s^{\prime}))=-k^{-1}\partial_{s}\cos(k(s-s^{\prime}))$.
Then, integrating over $k$ we obtain
$\partial_{s}\delta(s-s^{\prime})$ . In such a case the formula
for the density matrix in the limit $\beta\hbar \rightarrow 0$ is

\begin{equation}\begin{array}{l}
\rho_{\tau}({\bf X},{\bf X}^{\prime})\simeq\int {\cal D}X {\cal
D}X^{\prime}\exp\Big(\frac{i}{2\hbar}\int_{\tau_{0}}^{\tau}ds(\frac{dX_{r}}{ds}\frac{dX_{r}}{ds}
-\frac{dX_{r}^{\prime}}{ds}\frac{dX_{r}^{\prime}}{ds}) \cr
\exp\Big(-\frac{i}{2\pi\hbar}\int_{\tau_{0}}^{\tau}ds\Big((f-f^{\prime
})\partial_{s}(f+f^{\prime })
-\frac{1}{2\pi\hbar^{2}\beta}(f-f^{\prime })(f-f^{\prime
})\Big)\Big).\end{array}\end{equation} We expand eq.(143) around
$Q$.In the exponential (143) the term linear in $y$  becomes
\begin{equation}\begin{array}{l}
y_{n}\Big(-m\frac{d}{ds}\frac{dQ_{n}}{ds}+\frac{2\lambda^{2}m^{2}}{\pi}Q_{n}\frac{d^{5}}{ds^{5}}{\bf
Q}^{2}\Big).
\end{array}\end{equation}The contribution of the tensor
perturbations to the density matrix has been calculated in
\cite{habagrav}. It follows from eq.(140) with $f_{s}\rightarrow
f_{s}^{rl}$. After an expansion in $y$ the term linear in $y$
reads (we omit the contribution of classical solutions appearing
in eq.(86))

\begin{equation}\begin{array}{l}
y_{n}\Big(-m\frac{d}{ds}\frac{dQ_{n}}{ds}
+16Gm^{2}Q_{n}\frac{d^{5}}{ds^{5}}{\bf Q}^{2}+\frac{4
Gm^{2}}{5}Q_{l}\frac{d^{5}}{ds^{5}}(\frac{1}{3}Q_{r}Q_{r}\delta_{nl}-Q_{n}Q_{l})\Big)
\end{array}\end{equation}
The term quadratic in $y$ is the noise term. For low temperature
we obtain in general the non-local and non-Markovian stochastic
equation (115). In the high temperature limit $\beta\hbar
\rightarrow 0$ the calculation of the density matrix is reduced to
an expectation value over the solutions of the stochastic
differential equation
\begin{equation}\begin{array}{l}
-\frac{d^{2}Q_{n}}{ds^{2}}+16Gm Q_{n}\frac{d^{5}}{ds^{5}}{\bf
Q}^{2} +\frac{8\pi
Gm}{10\pi}Q^{l}\frac{d^{5}}{ds^{5}}(\frac{1}{3}Q_{r}Q_{r}\delta_{nl}-Q_{n}Q_{l})
\cr=m^{-1}\sqrt{2G}\beta^{-\frac{1}{2}}(M^{\frac{1}{2}})_{nr}\partial_{s}b_{s}^{r},
\end{array}\end{equation}(the tensor term on the
lhs of eq.(146) coincides with the one derived in \cite{wald}). As
explained in the derivation of eq.(115) the term quadratic in $y$
defines the operator $M$. From eq.(143) (after an insertion of
tensor perturbations) we obtain that $M$ is an operator defined by
the bilinear form
\begin{equation}\begin{array}{l}
2G\beta^{-1}y^{r}M^{rl}y^{l}=\frac{m^{2}\lambda^{2}}{4\pi}\beta^{-1}\int
ds\Big(
\frac{d^{2}}{ds^{2}}(Q^{j}y^{l})\frac{d^{2}}{ds^{2}}(Q^{j}y^{l})
\cr+\frac{d^{2}}{ds^{2}}(Q^{j}y^{l})\frac{d^{2}}{ds^{2}}(Q^{l}y^{j})
-\frac{2}{3}\frac{d^{2}}{ds^{2}}(Q^{j}y^{j})\frac{d^{2}}{ds^{2}}(Q^{l}y^{l})\Big)
\cr+8Gm^{2}\beta^{-1}\int ds \frac{d^{2}}{ds^{2}}y^{n}Q^{n}
\frac{d^{2}}{ds^{2}}y^{l}Q^{l}=
\frac{m^{2}\lambda^{2}}{4\pi}\beta^{-1}y^{r}Q^{k}{\cal
M}_{rk;ln}y^{l}Q^{n}
\end{array}\end{equation} where
\begin{equation}
{\cal M}_{rk;ln}(s,s^{\prime})=(<\frac{1}{4\pi}\Lambda_{rk;ln}>
+8\delta_{rk}\delta_{ln})\partial_{s}^{2}\partial_{s^{\prime}}^{2}\delta(s-s^{\prime})
\end{equation}
As in \cite{habagrav} we could derive eq.(146) together with the
noise (148) from the classical Gibbs distribution of gravitational
waves. The crucial check of the quantum nature of gravity would
involve a comparison of the experimental measurement of noise with
the truly quantum spectrum $k\coth(\frac{\hbar\beta k}{2}$
following from eq.(141).

In the radiation era ($a(\tau)=\tau$ ) there are minor changes in
our formula when applied to tensor perturbations. As follows from
eq.(116) the kinetic term changes into $\frac{1}{2}m\lambda
a(\frac{d{\bf X}}{ds})^{2}$ and $f^{rl}$ into
$a^{-1}\frac{d^{2}}{ds^{2}}aQ^{r}Q^{l}$  so that the tensor noise
term in eq.(147 ) is replaced by
\begin{equation}\begin{array}{l}
\frac{m^{2}\lambda^{2}}{4\pi}\beta^{-1}\int dsa^{-2}\Big(
\frac{d^{2}}{ds^{2}}(aQ^{j}y^{l})\frac{d^{2}}{ds^{2}}(aQ^{j}y^{l})\cr
+\frac{d^{2}}{ds^{2}}(aQ^{j}y^{l})\frac{d^{2}}{ds^{2}}(aQ^{l}y^{j})
-\frac{2}{3}\frac{d^{2}}{ds^{2}}(aQ^{j}y^{j})\frac{d^{2}}{ds^{2}}(aQ^{l}y^{l})\Big)
\end{array}\end{equation}
The lhs of eq.(146) resulting from tensor perturbations is changed
as
\begin{equation}\begin{array}{l}
-\frac{d}{ds}a\frac{dQ_{n}}{ds}+ \frac{8\pi
Gm}{10\pi}Q^{l}\frac{d^{3}}{ds^{3}}a^{-1}\frac{d^{2}}{ds^{2}}a(\frac{1}{3}Q_{r}Q_{r}\delta_{nl}-Q_{n}Q_{l})
\end{array}\end{equation} It follows that we can associate a
definite temperature to the tensor perturbations in the radiation
era. We obtain the radiation damping (150) and the noise (149)
which is proportional to the temperature $\beta^{-1}$. We cannot
do this for scalar perturbations. The contribution of the scalar
perturbations to the density matrix in the radiation era can be
treated by means of the methods of sec.13 with (from eq.(109))
$f_{s}({\bf k})=-m\lambda s^{-2}\frac{d^{2}}{ds^{2}}s{\bf X}^{2}$.
There is no thermal state for scalar perturbations in the
radiation era.  It is remarkable that the Hamiltonian (52) for the
scalar perturbations in the radiation era  is equal to the one for
the tensor perturbations (gravitons) (57) in the baryonic era
($a\simeq \tau^{2}$ when the velocity of light is replaced by the
acoustic velocity $c_{s}=\frac{1}{\sqrt{3}}$. For large conformal
time the term $a^{-1}a^{\prime\prime}\simeq 2\tau^{-2}$ is
negligible. This suggests that at the begin of the baryonic era (
recombination time) we could have the thermal state for
 tensor perturbations with the  temperature $T$ and the thermal state for
 scalar perturbations with the  temperature $T\sqrt{3}$.

  The spectrum of the noise following from eq.(141) is $8\pi G
  k\hbar\coth(\frac{\hbar\beta k}{2})$ which at low
  temperature is  $8\pi G
  k$ and at high temperature $8\pi G \beta^{-1}$.
\section{Discussion}
Our study is based on the quadratic approximations to the
Hamiltonian of tensor and scalar perturbations of Einstein
gravity. We explore the Schr\"odinger wave function as a solution
of the Schr\"odinger  equation in various epochs of universe
evolution. We are interested in the squeezing of the wave function
in various epochs. The squeezing of the wave function is relevant
for the motion of particles in the environment of cosmological
perturbations because it determines the intensity of the quantum
noise. We have calculated the density matrix resulting from an
average over quantized cosmological perturbations in a thermal
state and in a general Gaussian state. We have shown that the
evolution of the density matrix can be described by a stochastic
equation of radiation damping. The quantum modification of the
equation of motion can influence the formation of inhomogeneities
in the early stages of universe evolution. It can be seen that the
squeezing has little effect on the strength of the friction term
in radiation damping but can substantially enforce the noise term.
As earlier pointed out by Parikh,Wilczek and Zahariade the
quantized tensor perturbations could be detected in gravitational
wave detectors as a specific noise. We have shown in this paper
that this noise is modified by scalar perturbations. The
determination of the  exact form of the quantum noise in detectors
may be important for distinguishing it from other sources of
noise.Besides the gravitational wave detectors the scalar
cosmological perturbations should have an impact on primordial
black holes formation, CMB temperature fluctuations and on
galaxies distribution.

   In  view of the prospective development of the detection of
   cosmological perturbations it is interesting to extend our
   method to modified theories of gravity and non-perturbative
   quantizations. The question arises whether the quadratic
   approximation to the Hamiltonian and Gaussian approximation
   to the wave function of the gravitational perturbations applies
   in these theories. If so then the treatment of sec.10 of the
   interaction of gravitational perturbations with a
   non-relativistic particle (detector) can be used in order to
   obtain a stochastic equation for the radiation damping.
   Some of the modified theories of gravity (Horndeski, $P(X,\phi)$)
   predict the speed of propagation of gravitational waves
   different from the speed of light. This (hypothetical)
   difference is carefully studied in present day observations.
    The results on the particle
   interaction with cosmological perturbations in such theories
   could further  be used for a selection of a proper model on the
   basis of detection experiments.
We did not insert numerical values in our mathematical results.
Numerical estimates for solutions of the evolution equations are
needed for a comparison with  observations. Such detailed
numerical studies are postponed to a prospective research.


\begin{thebibliography}{99}\bibitem{bert1}E.Bertschinger,
Simulation of structure formation in the universe, Annual
Rev.Astronomy Astroph.{\bf 36},599(1998)
\bibitem{bert2}E.Bertschinger, Cosmological perturbation theory
and structure formation, arXiv:astro-ph/9408028,1994
\bibitem{star1}A.Starobinsky,Dynamics of phase transition in the
new inflationary universe, Phys.Lett.{\bf B117},175(1982)
\bibitem{guth}A. Guth and S.-Y. Pi, Quantum mechanics of the
scalar field in the new inflationary universe,Phys.Rev.{\bf
D32},679(1985)
\bibitem{hawking}S. Hawking, The development of irregularities in
a single bubble inflationary universe,Phys.Lett.{\bf
B115},295(1982)
\bibitem{chibisov}V.M. Mukhanov and G.V. Chibisov, Vacuum energy
and large-scale structure of the universe, Soviet Phys.JETP{\bf
56},258(1982)


\bibitem{bing} S. Boughn and T. Rothman,Aspects of graviton
detection:graviton emission and absorption by atomic hydrogen,

 Class.Quant.Grav.{\bf
23},5839(2006)
\bibitem{guer}T. Guerreiro, Quantum effects in gravity waves,

Class.Quant.Grav.{\bf 37},155001(2020)
\bibitem{bardeen}J. Bardeen, Gauge-invariant
cosmological perturbations,

Phys.Rev.{\bf D22},1882(1980)
\bibitem{mukhanov}V.F. Mukhanov, H.A. Feldman and R.H.
Brandenberger, Theory of cosmological perturbations, Phys.Reports,
{\bf 215},203(1992)
\bibitem{mukhanovbook}V. Mukhanov, Physical Foundations of
Cosmology, Cambridge,2005



\bibitem{sakai}M. Sasaki, Large scale quantum fluctuations in the inflationary universe, Progr.Theor.Phys.{\bf 76},1036(1986)

\bibitem{grishchuk}L.P. Grishchuk and Y.V. Sidorov,
Squeezed quantum states of relic gravitons and primordial density
fluctuations, Phys.Rev.{\bf D42},3413(1990)



\bibitem{aa}A.Albrecht,P.Ferreira,M.Joyce and T. Prokopec,
Inflation and squeezed quantum states, Phys.Rev.{\bf
D50},4807(1994)
\bibitem{star}D. Polarski and A.A. Starobinsky,Semiclassicality
and decoherence of cosmological perturbations,
 Class.Quantum Grav.{\bf 13},377(1996)


\bibitem{starpol}J. Lesgourgues, D. Polarski and A.A. Starobinsky,
Quantum-to-classical cosmological perturbations for non-vacuum
initial states, Nucl.Phys.{\bf B497},479(1997)
\bibitem{wilczek}M. Parikh, F. Wilczek and G.
Zahariade,The noise of gravitons,

Int.Journ.Mod.Phys {\bf D29},2042001(2020),



doi.org/10.1142/S0218271820420018,
 arXiv:2005.07211[hep-th]
 \bibitem{wilczek1}M. Parikh, F. Wilczek and G.
Zahariade, Quantum mechanics of gravitational waves,
arXiv:2010.08205[hep-th]
\bibitem{wilczek2}M. Parikh, F. Wilczek and G.
Zahariade, Signatures of the quantization of gravity at
gravitational wave detectors, arXiv:2010.08208[hep-th]

\bibitem{habagrav}Z.Haba, State-dependent graviton noise in the equation of geodesic deviation,
Eur.Phys.J. {\bf C81},40(2021)
\bibitem{soda} S. Kanno, J.Soda and J. Tokuda, Noise and
decoherence induced by gravitons, Phys.Rev.{\bf D103},044017(2021)


\bibitem{hu}E. Calzetta and B.L. Hu, Noise and fluctuations in semiclassical gravity, Phys.Rev.{\bf D49}, 6636
(1994)

\bibitem{ango}C. Anastopoulos, Quantum theory of non-relativistic
particles interacting with gravity, Phys.Rev.{\bf D54},1600(1996)

 \bibitem{habarel}Z. Haba, Decoherence by relic gravitons,Mod.Phys.Lett.{\bf
A15},1519(2000)
\bibitem{hk} Z.Haba and H. Kleinert, Quantum Liouville and
Langevin equations for gravitational radiation damping,

Int.J.Mod.Phys.{\bf A17},3729(2002),arXiv:quant-ph/0101006
\bibitem{angohu}C. Anastopoulos and B.L.Hu,
A master equation for gravitational decoherence:probing the
textures of spacetime,

Class.Quant.Gravity {\bf 30}, 165007 (2013)

\bibitem{wang} T. Oniga and Ch.H.-T.Wang,
Quantum gravitational decoherence of light and matter,
Phys.Rev.{\bf D93},044027(2016)



 \bibitem{bao}
Tao Hong, J.L.Han  and Z.L. Wen, A detection of Baryon Acoustic
Oscillations from the distribution of galaxy clusters,
Astroph.J.{\bf 826},154(2016)
\bibitem{tensor1}K.N. Ananda,Ch. Clarkson and D.Wands,


Cosmological gravitational wave background from primordial density
perturbations, Phys.Rev.{\bf D75},123518(2007)
\bibitem{tensor2} Chen Yuan and Qing-Guo Huang, A topic review on
probing primordial black hole dark matter with scalar induced
gravitational waves, arXiv:[astro-ph.GA]2103.04739, 2021
\bibitem{allen}B. Allen,E.E. Flanagan and M.A. Papa, Is the
squeezing of relic gravitational waves produced by inflation
detectable? Phys.Rev.{\bf D61},024024,1999
\bibitem{salam}C. J. Isham, Abdus Salam and J.S. Strathdee,
Infinity supression in gravity-modified
electrodynamics.II,Phys.Rev.{\bf D5},2548(1972)
\bibitem{habaarx}Z. Haba, Renormalization in quantum Brans-Dicke gravity,
hep-th/0205130,2002
\bibitem{russ}G. Vereshkov and L. Marochnik, Quantum gravity in
Heisenberg representation and self-consistent theory of gravitons
in macroscopic space-time, J.Mod.Phys.{\bf 4},285(2013)
\bibitem{ashtekar} A. Ashtekar, C. Rovelli and L. Smolin,
Gravitons and loops, Phys. Rev.{\bf D44},1740(1991)
\bibitem{effective}J.F. Donoghue, Introduction to the effective field theory description of gravity,
arXiv:gr-qc/9512024,1995
\bibitem{safe}A. Bonnano, T. Denz, J.M.Pawlowski and M.Reichert,
Reconstructing the graviton, arXiv:hep-th/2102.02217
\bibitem{odintsov}S. Nojiri and S.D. Odintsov, Unified cosmic history in modified gravity:
from f(R) theory to Lorentz non-invariant models, Physics Reports
{|bf 505},59,2011\bibitem{brans}C.H. Brans and R.H. Dicke,Mach's
principle and a relativistic theory of gravitation, Phys.Rev.{\bf
124},925(1961)
\bibitem{horndeski}
E. Belgacem et al, Testing modified gravity at cosmological
distances with LISA standard sirens,
arXiv:astro-ph.CO/1906.01593\bibitem{fer}P.G. Ferreira,
Cosmological tests of gravity,arXiv:1902.10503,2019

\bibitem{velocity}D. Bettoni,J.M. Ezquiaga, K. Hinterbichler and
M.Zumalcarregui, Speed of gravitational waves and the fate of
scalar-tensor gravity, Phys.Rev.Lett.{\bf D95},084029(2017)
\bibitem{freidlin}M. Freidlin, Functional Integration and Partial
Differential Equations, Princeton, Univ.Press,1995

\bibitem{simon} B.Simon, Functional Integration and Quantum
Physics, Academic Press, New York,1979





\bibitem{habajpa}

Z. Haba, Feynman integral and complex classical trajectories,
Lett.Math.Phys.{\bf 37},223(1996)
\bibitem{hababook}Z. Haba, Feynman Integral and Random Dynamics in
Quantum Physics,Kluwer/Springer,1999

\bibitem{doss1}H. Doss, On a stochastic solution of the Schroedinger equation with analytic coefficients  Commun.Math.Phys.{\bf 73},247(1980)
\bibitem{alb} S.Albeverio, Z. Brzezniak and Z. Haba,
 On the Schr\"odinger equation with potentials which are Laplace
 transforms of measures,
Potential Analysis, {\bf 9},65 (1998)

\bibitem{doss2} H. Doss, On a probabilistic approach to the Schr\"odinger equation
with a time-dependent potential,Journ.Funct.Anal.{\bf
260},1824(2011)

\bibitem{bassett}B.A.  Bassett, S. Tsujikawa and D. Wands,
 Inflaton dynamics and reheating, Rev.Mod.Phys.{\bf 78},537(2006)


\bibitem{schwarz}J. Martin and D.J. Schwarz,
Precision of slow-roll prediction for cosmic microwave background
radiation anisotropies, Phys.Rev.{\bf D62},103520(2000)

\bibitem{warmfriction}F. D'Eramo and K.Schmitz, Imprint of scalar
era on the primordial spectrum of gravitational waves,
Phys.Rev.Research.{\bf 1},013010(2019)

\bibitem{bunch}T.S.Bunch  and P.C.W. Davis, Quantum field theory in de Sitter space: renormalization by point splitting,
Proc.Roy.Soc.London, {\bf 360},117(1978)


\bibitem{wise}L.F. Abbott and M.B. Wise, Constraints on
generalized inflationary cosmologies, Nucl.Phys.{\bf
B244},541(1984)






\bibitem{zurek}K.M. Zurek, On vacuum fluctuations in quantum gravity and interferometer arm fluctuations,  arXiv:2012.05870[hep-th]


\bibitem{benguria}R. Benguria and M. Kac, Quantum Langevin equation, Phys.Rev.Lett.
{\bf 46},1(1981)


\bibitem{ford}G.W. Ford, J.T. Lewis and R.F. O'Connell, Quantum
Langevin equation, Phys.Rev. {\bf A37},4419(1988)

\bibitem{feynman}R.P.Feynman and F.L. Vernon, The theory of general quantum system
interacting with a linear dissipative system, Annals Phys.{\bf
24},118 (1963)



\bibitem{calchu}


B.L. Hu, J.P. Paz and Y.H. Zhang, Quantum Brownian motion in a
general environment: Exact master equation with nonlocal
dissipation and colored noise, Phys.Rev.{\bf D45},2843(1992)






\bibitem{kleinert}H. Kleinert, Path Integrals, 5th edition, World
Scientific, 2009



\bibitem{wald}T.C. Quinn and R.M.  Wald,  Axiomatic approach to electromagnetic
 and gravitational radiation reaction of particles in curved space-time,
Phys.Rev.{\bf
D56},3381(1997)



\end{thebibliography}
\end{document}